\documentclass[10pt,draftcls, onecolumn, journal]{IEEEtran}

\usepackage{amsmath,
amssymb}
\usepackage{stmaryrd}
\usepackage{epsfig,calc}
\usepackage{wasysym}

\title{Dictionary Identification -\\
Sparse Matrix-Factorisation via $\ell_1$-Minimisation}

\author{R{\'e}mi Gribonval, {\em Senior Member, IEEE}, \and Karin Schnass
\thanks{This work was supported in part by the European Commission through the SMALL project under FET-Open grant number: 225913.}
\thanks{R{\'e}mi Gribonval is with Projet METISS, 
Centre de Recherche INRIA Rennes - Bretagne Atlantique, 
IRISA, Campus de Beaulieu, 
F-35042 Rennes Cedex, France, 
E-mail: firstname.lastname@irisa.fr}
\thanks{Karin Schnass is with the Johann Radon Institute for Computational and Applied Mathematics (RICAM), Altenbergerstra\ss e 54, 4040 Linz, Austria,
	E-mail: firstname.lastname@oeaw.ac.at }
}

\newcommand\dico{\mathbf{\Phi}}
\newcommand\atom{\varphi}

\newcommand\ip[2]{\langle #1, #2\rangle}
\newcommand\natoms{K}
\newcommand\sparsity{S}
\newcommand\ddim{d}
\newcommand\good{\Lambda}
\newcommand\bad{{\overline{\good}}}

\newcommand\Id{\mathbf{I}}

\newcommand\ie{{i.e. }}
\newcommand\iid{{i.i.d. }}
\newcommand\Y{Y}
\newcommand\X{X}

\newcommand\eps{\varepsilon}
\newcommand\manifold{\mathcal{D}}
\newcommand\tangentspace[2]{T_{#2}#1}

\newcommand\pnorm{{q}}
\newcommand\randsign[0]{\xi}
\def\radius{R}

\def \M {\mathbf{M}_0}

\def \V {\mathbf{V}}
\def \Zdiag {\mathbf{Z}}
\def \Diag {\mathbf{\Delta}}

\newcommand\nsig[0]{N}

\newcommand\Cost[1]{\mathcal{C}_{#1}}
\newcommand\Null{\mathcal{N}}

\newcommand\epscover{\mathcal{X}}

\newcommand\diag{\operatorname{diag}}

\newcommand\sign{\operatorname{sign}}
\newcommand\trace{\operatorname{trace}}

\newcommand{\R}{{\mathbb{R}}}

\renewcommand{\P}{{\mathbb{P}}}
\newcommand{\E}{{\mathbb{E}}}

\newtheorem{Theorem}{Theorem}[section]
\newtheorem{lemma}[Theorem]{Lemma}
\newtheorem{corollary}[Theorem]{Corollary}

\newtheorem{remark}{Remark}[section]

\newenvironment{Proof}{\noindent
{\bf\underline{Proof:} }}
{\hspace*{\fill}$\Box$\vskip1em}

\begin{document}

\maketitle

\begin{abstract}
This article treats the problem of learning a dictionary providing sparse representations for a given signal class, via $\ell_1$-minimisation. The problem can also be seen as factorising a $\ddim \times \nsig$ matrix $Y=(y_1 \ldots y_\nsig), \, y_n\in \R^\ddim$ of training signals into a $\ddim \times \natoms$ dictionary matrix $\dico$ and a $\natoms \times \nsig$ coefficient matrix  $\X=(x_1\ldots x_\nsig),\, x_n \in \R^\natoms$, which is sparse. The exact question studied here is when a dictionary coefficient pair $(\dico,\X)$ can be recovered as local minimum of a (nonconvex) $\ell_1$-criterion with input $Y=\dico \X$. First, for general dictionaries and coefficient matrices, algebraic conditions ensuring local identifiability  are derived, which are then specialised to the case when the dictionary is a basis. Finally, assuming a random Bernoulli-Gaussian sparse model on the coefficient matrix, it is shown that sufficiently incoherent bases are locally identifiable with high probability. The perhaps surprising result is that the typically sufficient number of training samples $\nsig$ grows up to a logarithmic factor only linearly with the signal dimension, i.e. $\nsig \approx C \natoms \log \natoms$, in contrast to previous approaches requiring combinatorially many samples. 
\end{abstract}

\begin{keywords}
$\ell_1$-minimisation, compressed sensing, random matrices, sparse representation, dictionary learning, dictionary identification, nonconvex optimisation, independent component analysis, blind source separation, blind source localisation.
\end{keywords}

\section{Introduction} \label{sec:intro}

Many signal processing tasks, such as denoising and compression, can be efficiently performed if one knows a sparse representation of the signals of interest. Moreover, a huge body of recent results on sparse representations has highlighted their impact on inverse linear problems such as (blind) source separation and localisation as well as compressed sampling, for a starting point see e.g. \cite{Tropp:greed, bp:fuchs, donoho:bp, tr06}. 
\\
In any of these publications, one will - more likely than not - find a statement starting with 'given a dictionary $\dico$ and a signal $y$ having an $\sparsity$-sparse approximation/representation $y = \dico x$ \ldots', which points exactly to the remaining problem: all applications of sparse representations rely on a signal dictionary $\dico$ from which sparse linear expansions can be built that efficiently approximate the signals from a class of interest; success heavily depends on the good fit between the data class and the dictionary. 
\\
For many signal classes, good dictionaries -- such as time-frequency or time-scale dictionaries -- are known, but new data classes may require the construction of new dictionaries to fit new types of data features. The analytic construction of dictionaries such as wavelets and curvelets stems from deep  mathematical tools from Harmonic Analysis. It may, however, be difficult and time consuming to develop complex mathematical theory each time a new class of data, which requires a different type of dictionary,  is met. An alternative approach is dictionary learning, which aims at infering the dictionary $\dico$ from a set of training data $y_n$. Dictionary learning, also known as {\em sparse coding}, has the potential of 'industrialising' sparse representation techniques for new data classes. 
\\
This article treats the theoretical dictionary learning problem, expressed as a factorisation problem which consists of identifying a $\ddim \times \natoms$ matrix $\dico$ from a set of $\nsig$ observed training vectors $y_n\in \R^\ddim$,  knowing that $y_n = \dico x_n$, $1 \leq n \leq N$ for some unknown collection of coefficient vectors $x_n\in \R^\natoms$ with certain statistical properties. 
\\
Considering the extensive literature available for the sparse decomposition problem after the early work in ~\cite{dohu01,grini03,donoho:bp,carota05,Tropp:relax}, surprisingly little work has been dedicated to theoretical dictionary learning so far. 
There exist several dictionary learning algorithms (see e.g. \cite{olsfield96, kreutz03, ahelbr06, Jost:motif}), but only recently people have started to consider also the theoretical aspects of the problem. The origins of research into what is now called dictionary learning can be found in the field of Independent Component Analysis (ICA)~\cite{Com94,cardoso98}. There, many identifiability results are available, which, however, rely on {\em asymptotic} statistical properties under {\em statistical independence} and {\em non-Gaussianity} assumptions. 
\\
In contrast, Georgiev, Theis and Cichocki, \cite{gethci05}, as well as Aharon, Elad and Bruckstein, \cite{ahelbr06b}, described more geometric identifiability conditions on the sparse coefficients of training data in an ideal (overcomplete) dictionary. 
Yet, for these conditions to hold, the size $\nsig$ of the training set seems to be required to grow exponentially fast with the number of atoms $\natoms$, and the provably good identification algorithms are combinatorial. Moreover, the algorithms and the identifiability analysis are not robust to 'outliers', i.e., training samples $y_n$ where $x_n$ fails to be sufficiently sparse.
For applications, on the other hand, we are concerned with relatively large-dimensional data (e.g. $\ddim= 30$, or even $\ddim = 1000$) but limited availability of training data ($\nsig$ is not much larger than say $1000 \cdot \ddim$) as well as limited computational resources. \\
In this article, we study the possibility of designing provably good, non-combinatorial dictionary learning algorithms that are robust to outliers and to the limited availability of training samples. Inspired by recent proofs of good properties of $\ell_1$-minimisation for sparse signal decomposition with a given dictionary, we investigate the properties of $\ell_1$-based dictionary learning, \cite{zipe01,pl07}. Our ultimate goal, described in details in Section~\ref{sec:setting}, is to characterise properties that a set of training samples $y_n, 1 \leq n \leq \nsig$ should satisfy to guarantee that an ideal dictionary is the only local minimum of the $\ell_1$-criterion, opening up the possibility of replacing combinatorial learning algorithms with efficient numerical descent techniques.  As a first step, we investigate conditions under which an ideal dictionary is a local minimum of the $\ell_1$-criterion.
\\
{\bf Main results.} 
First,  we describe the proposed setting in Section~\ref{sec:setting} and characterise the local minima of the $\ell_1$-cost function in Section~\ref{sec:LocalIdentifiability}. We discuss the geometrical interpretation of this characterisation in Section~\ref{sec:GeometricInterpretation}. Then, using concentration of measure, we prove in Section~\ref{sec:probana} the perhaps surprising result that when  
\[
\nsig \geq C K \log K,
\]
if the samples $x_n, 1 \leq n \leq \nsig$, are a typical draw from a Bernoulli-Gaussian random distribution (which can generate a large proportion of {\em outliers}), then any sufficiently incoherent basis matrix $\dico$, \ie $\natoms =\ddim$, is a local minimum of the cost function and is therefore 'locally identifiable'. The constant $C$ depends on a parameter of the Bernoulli-Gaussian distribution which drives the sparsity of the training set.\\
This number of training samples is surprisingly small considering that $\nsig$ training samples provide $\nsig \times \natoms \geq C \natoms^2 \log \natoms$ real parameters, while the basis matrix $\dico$ is essentially parameterised by $O(\natoms^2)$ independent real parameters.\\
In the considered matrix identification setting, it should be noted that $\ell_1$ is {\em not} a convex cost function. It admits {\em several local minima} hence local identifiability only implies that, upon good initial conditions, numerical optimisation schemes performing the $\ell_1$-optimisation will recover the desired matrix $\dico$. However, empirical experiments in low dimension ($\ddim=2$), shown in Section~\ref{sec:discussion}, indicate that for typical draws of Bernoulli-Gaussian training samples $x_n$, the matrix $\dico$ is in fact the {\em only} local minimum of the criterion (up to natural indeterminacies of the problem such as column permutation). If this empirical observation could be turned into a theorem for general dimension $K$ under the Bernoulli-Gaussian sparse model, this would imply that typically: a) $\ell_1$-minimisation is a good {\em identification principle}; b) any decent $\ell_1$-descent algorithm is a good {\em identification algorithm }.

\section{Setting \label{sec:setting}} 
\noindent In the vector space $\mathcal{H} = \R^\ddim$ of $\ddim$-dimensional signals, a dictionary is a collection of $\natoms \geq \ddim$ 
vectors $\atom_k$, $1 \leq k \leq \natoms$, and it is said to be {\em complete} if its columns span the whole space.  Alternatively, a dictionary can be seen as a $\ddim \times \natoms$ matrix $\dico$. 
For a given signal $y \in \mathcal{H}$, the sparse representation problem consists of finding a representation $y = \dico \cdot x$ where $x \in \R^\natoms$ is a 'sparse' vector, {\em i.e.} with  few significantly large coefficients and most of its coefficients negligible. 

\subsection{Sparse Representation by $\ell_1$-Minimisation, with a Known Dictionary}
\noindent For a given dictionary, selecting  an 'ideal' sparse representation of some data vector $y \in \mathcal{H}$ amounts to solving the problem
\begin{equation}
\label{eq:P0}
\min_{x} \|x\|_0,\ \mbox{such that}\ \dico x = y
\end{equation}
where the $\ell_0$ pseudo-norm $\|x\|_0$ counts the number of nonzero entries in the vector $x$. However, being nonconvex and nonsmooth,~\eqref{eq:P0} is hard to solve and has indeed been shown to be an NP-hard problem~\cite{davis:mp,na95}. As a result people turned to non optimal strategies like greedy algorithms or the Basis Pursuit Principle. There the problem above is replaced by its convex relaxation
\begin{equation}
\label{eq:P1}
\min_{x} \|x\|_1,\ \mbox{such that}\ \dico x = y.
\end{equation}
The good news is that when $y$ admits a sufficiently sparse representation the solution of the relaxed problem coincides with the solution of the original one, compare \cite{grini03, donoho:bp, carota05, Tropp:relax}.

\subsection{Dictionary Learning from a Collection of Training Samples}
\noindent A related problem is that of finding the dictionary that will fit a class of signals, in the sense that it will  provide sparse representations for all signals of the class. The first idea is to find the dictionary allowing representations with the most zero coefficients, \ie given $\nsig$ signals $y_n \in \mathcal{H}$, $1 \leq n \leq \nsig$, and a candidate dictionary $\dico$, one can measure the global sparsity as
\[
\sum_{n=1}^\nsig \min_{x_n} \|x_n\|_0,\ \mbox{such that}\  \dico x_n = y_n, \, \forall n.
\] 
Collecting all signals $y_n$ (considered as column vectors in $\R^\ddim$) into a $\ddim \times \nsig$ matrix $\Y$ and all coefficients $x_n$ (considered as column vectors in $\R^\natoms$) into a $\natoms \times \nsig$ matrix $\X$, the fit between a dictionary $\dico$ and the training signals $\Y$ can be measured by the cost function
\[
\Cost{0}(\dico|\Y) := \min_{\X \ |\ \dico \X = \Y} \|\X\|_0,
\]
where 
$
\|\X\|_0 := \sum_n \|x_n\|_0
$
counts the total number of nonzero entries in the $\natoms \times \nsig$ matrix $\X$. Thus to get the dictionary providing the most zero coefficients out of a prescribed collection $\manifold$ of admissible dictionaries, we should consider the criterion
\begin{equation} \label{eq:LearningL0}
\min_{\dico \in \manifold} \Cost{0}(\dico|\Y).\tag{P0}
\end{equation}

The problem is that already finding the representation with minimal non-zero coefficients for one signal in a given dictionary is np-hard, which makes trying to solve \eqref{eq:LearningL0} indeed a daunting task. 
Fortunately the problem above is not only daunting but also rather uninteresting, since it is not stable with respect to noise or suited to handle signals that are only compressible.
Thus the idea of learning a dictionary via $\ell_1$-minimisation is motivated on the one hand by the goal to have a criterion that is taking into account that the signals might be noisy or only compressible and on the other by the success of the Basis Pursuit principle for finding sparse representations. There the $\ell_0$-pseudo norm was replaced with the $\ell_1$-norm, which also promotes sparsity but is convex and continuous. The same strategy can be applied to the dictionary learning problem and the $\ell_0$-cost function can be replaced with the $\ell_1$-cost function 
\begin{equation} \label{eq:LearningCost}
\Cost{1}(\dico|\Y) := \min_{\X \ |\ \dico \X = \Y} \|\X\|_1,
\end{equation}
where $\|\X\|_1 := \sum_n \|x_n\|_1$. 
Several authors, \cite{zipe01, kreutz03, pl05, Peol06, pl07, yablda08, hlrata09}, have proposed to consider the corresponding minimisation problem
\begin{equation} \label{eq:Learning}
\min_{\dico \in \manifold} \Cost{1}(\dico|\Y).\tag{P1}
\end{equation}
Unlike for the sparse representation problem, where this change meant a convex relaxation, the dictionary learning problem~\eqref{eq:Learning} is still {\em not convex} and cannot be immediately addressed with generic convex programming algorithms\footnote{The problem investigated here should not be confused with the problem of sparse channel estimation considered by 
Pfander, Rauhut and Tanner in~\cite{pfrata08}. There the goal is to identify a transmission channel $\dico$ by an appropriate choice of input sequence $x$ and the observation of $y = \dico x$. The approach is to model $\dico = \sum_\ell \alpha_\ell \dico_\ell$ with sparse coefficients $\alpha$ in a {\em known} dictionary of "atomic channels", and to solve the convex problem $\min \|\alpha\|_1$ subject to $y = \sum_\ell \alpha_\ell (\dico_\ell x)$. Here, we do not have the freedom to choose $x$ nor do we know the channel dictionary, and the problem we consider is no longer convex.}. However, it seems better behaved than the original problem~\eqref{eq:LearningL0} because of the continuity of the criterion with respect to increasing amounts of noise, which makes it more amenable to numerical implementation. \\
Looking at the problem above, we see that in order to solve it we still need to define $\manifold$, the set of admissible dictionaries. 

\subsection{Constraints on the Dictionary}
\noindent Several families of admissible dictionaries can be considered such as discrete libraries of orthonormal bases (wavelet packets or cosine packets, for which fast dictionary selection is possible using tree-based searches \cite{cowi92}). Here we focus on the 'non parametric' learning problem where the full $\ddim \times \natoms$ matrix $\dico$ has to be learned.
Since the value of the criterion in~\eqref{eq:Learning} can always be decreased by jointly replacing $\dico$ and $\X$ with $\alpha \dico$ and $\X/\alpha$, $0 < \alpha < 1$, a scaling constraint is necessary and a common approach is to only search for the optimum of~\eqref{eq:Learning} within a bounded domain $\manifold$. \\
We propose to concentrate on inequality constraints of the form\footnote{Other constraints which replace the norm $\|\atom_k\|_2$ with, e.g., a norm $\|\atom_k\|_1$, would also be interesting to study when it is desirable to obtain sparse atoms and not only sparse coefficients.}
$
\max_k \|\atom_k\|_2 \leq C.
$
Because of the homogeneity of the criterion with respect to scaling, we can assume without loss of generality that $C=1$. We also let the reader check that the optimum of~\eqref{eq:Learning} with 
the considered inequality constraints is indeed achieved when there is equality, see also ~\cite{kreutz03,yablda08}. Hence we define the following constraint manifold
\begin{equation}
\label{eq:DicoConstraintUnit}
\manifold
 := \{\dico, \forall k, \|\atom_k\|_2 = 1\}.
\end{equation}
Let us turn now to the special aspect of dictionary learning treated in this paper.

\subsection{Dictionary Recovery: the Identification Problem}
\noindent Several algorithms have been proposed which adopt an $\ell_1$ minimisation approach to learning a dictionary, \cite{olsfield96, kreutz03,pl07}, from training data. Their empirical behaviour has been explored, showing their ability to often recover with good precision the underlying dictionary.

Here we are interested in the more theoretical problem of {\em dictionary identification} by $\ell_1$-minimisation: assuming that the data $\Y$ were generated from an 'ideal' dictionary $\dico_0 \in \manifold$ and 'ideal' coefficients $\X_0$ as $\Y = \dico_0 \X_0$, we want to determine conditions on $\X_0$ (and to a lesser extent on $\dico_0$) such that the minimisation of~\eqref{eq:Learning} recovers $\dico_0$. Our objective is therefore similar in spirit to previous work on dictionary recovery
\cite{gethci05,ahelbr06b} which studied the uniqueness of overcomplete dictionaries for sparse component analysis. The main difference here  is that we specify in advance which optimisation criterion we want to use to recover the dictionary ($\ell_1$-minimisation) and attempt to express conditions on a matrix $\X_0$ to guarantee that this method will successfully  recover a given class of dictionaries. \\
\noindent {\bf Permutation and sign ambiguity.} The first problem we face consists of the ambiguities, which have been well known since the development of ICA. Because of the normalisation constraint we are assuming on the dictionary, the usual scaling ambiguity is avoided, but there remains a permutation and a sign ambiguity: for any permutation matrix ${\bf P}$ and ${\bf D}$ any diagonal matrix with unit diagonal entries we have  $\dico \X = (\dico {\bf P^{-1} D^{-1} }) ({\bf DP} \X)$. Hence Problem~\eqref{eq:Learning} has not just one but a whole equivalence class of minimisers, each of them corresponding to a matching column resp. row permutation and sign change of $\dico$ resp.  $\X$. Therefore, we have to relax our requirement and only ask to find conditions such that minimising~\eqref{eq:Learning} recovers $\dico_0$ up to permutation and sign change.  The notation $\dico \sim \dico_0$ will indicate this indeterminacy, meaning that $\dico = \dico_0 {\bf P D}$ for some permutation matrix ${\bf P}$ and diagonal matrix ${\bf D}$ with unit diagonal entries.\\
\noindent {\bf Global identifiability {\em vs} local identifiability.} Ideally, we would like to characterise coefficient matrices $\X_0$ such that, for any $\dico_0 \in \manifold$ (or at least for a reasonable subset of $\manifold$ such as, for instance, 'incoherent' dictionaries), the {\em global minima} of 
\begin{equation}
\label{eq:Learning1}
\min_{\dico \in \manifold} \Cost{1}(\dico|\dico_0 \X_0) 
\end{equation}
can only be found at $\dico \sim \dico_0$.\\
An even more ambitious objective would be to characterise coefficient matrices such that the {\em  local minima} of~\eqref{eq:Learning1} can only be found at $\dico \sim \dico_0$, which would guarantee that  numerical optimisation algorithms  cannot be trapped in spurious local minima, and would converge independently of their initialisation.  This objective raises two complementary questions:
\renewcommand {\theenumi}{\alph{enumi}}
\begin{enumerate}
\item {\em Local identifiability:} Which conditions on $\X_0$ (and $\dico_0$) guarantee that  $\dico_0$ is a {\em local minimum} of the $\ell_1$-cost function? 
\item {\em Uniqueness:}  Which conditions guarantee that, when $\dico$ is a local minimum of the $\ell_1$-cost function, it must match $\dico_0$ up to column permutation and sign change?
\end{enumerate}
In this paper we concentrate on the first question. The characterisation of local minima of the $\ell_1$ criterion that we carry out in Section~\ref{sec:LocalIdentifiability} will certainly serve to address the second question in future work.\\

\noindent{\bf Ideally sparse training samples {\em vs} non-sparse outliers}
In contrast to previous theoretical work on dictionary uniqueness \cite{gethci05,ahelbr06b}, we wish to determine identification conditions that do not rely on the unrealistic assumption that each training sample is ideally sparse. As a first step to deal with training data which may contain training samples 
$y_n = \dico_0 x_n$ with non-sparse coefficients $x_n$,  we consider in Section~\ref{sec:probana} a Bernoulli-Gaussian model and show that, when the number of training samples drawn according to this model is sufficiently high, incoherent bases are associated to local minima of~\eqref{eq:Learning1}.

\begin{figure}[htbp]
\begin{center}
\includegraphics[width=\textwidth/2]{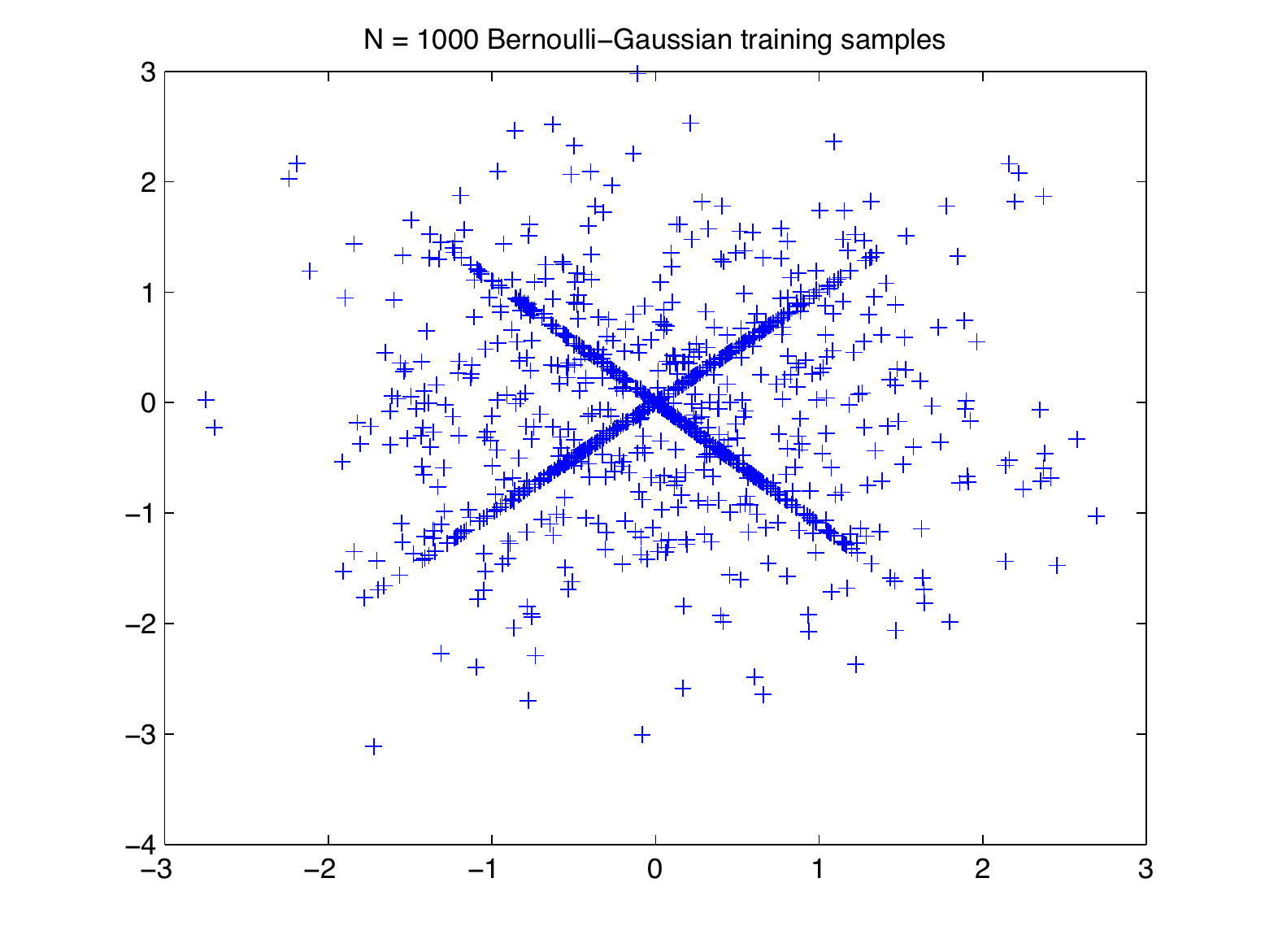}
\caption{A cloud of $\nsig = 1000$ training samples in $\R^2$. Each point is a column $y_n$ of the matrix $\Y = \dico_0 \X_0$, where $\X_0$ was generated using the Bernoulli-Gaussian model of Section~\ref{sec:probana} with $p = 0.7$.}
\label{fig:cloudBG}
\end{center}
\end{figure}

Figure~\ref{fig:cloudBG} illustrates a typical cloud of $\nsig = 1000$ points $y_n  = \dico_0 x_n \in \R^\ddim$, $\ddim = 2$, where $x_n$ was generated according to this Bernoulli-Gaussian model with parameter $p = 0.7$ (cf Section~\ref{sec:probana}). Here the dictionary is a basis made of two atoms $\atom_k^\star = (\cos \theta_k^\star, \sin \theta_k^\star)^T \in \R^2$, $k=0,1$, characterised by their angle $\theta_k^\star$, and its coherence is $\mu = |\langle \atom_0^\star,\atom_1^\star\rangle| = |\cos (\theta_1^\star-\theta_0^\star)| = 0.05$. One can observe that, while many training samples are perfectly aligned with the lines generated by the two atoms of the dictionary, there is also a substantial proportion of "outliers" that do not have a sparse representation in the considered dictionary.

\begin{figure}[htbp]
\begin{center}
\includegraphics[width=\textwidth/2]{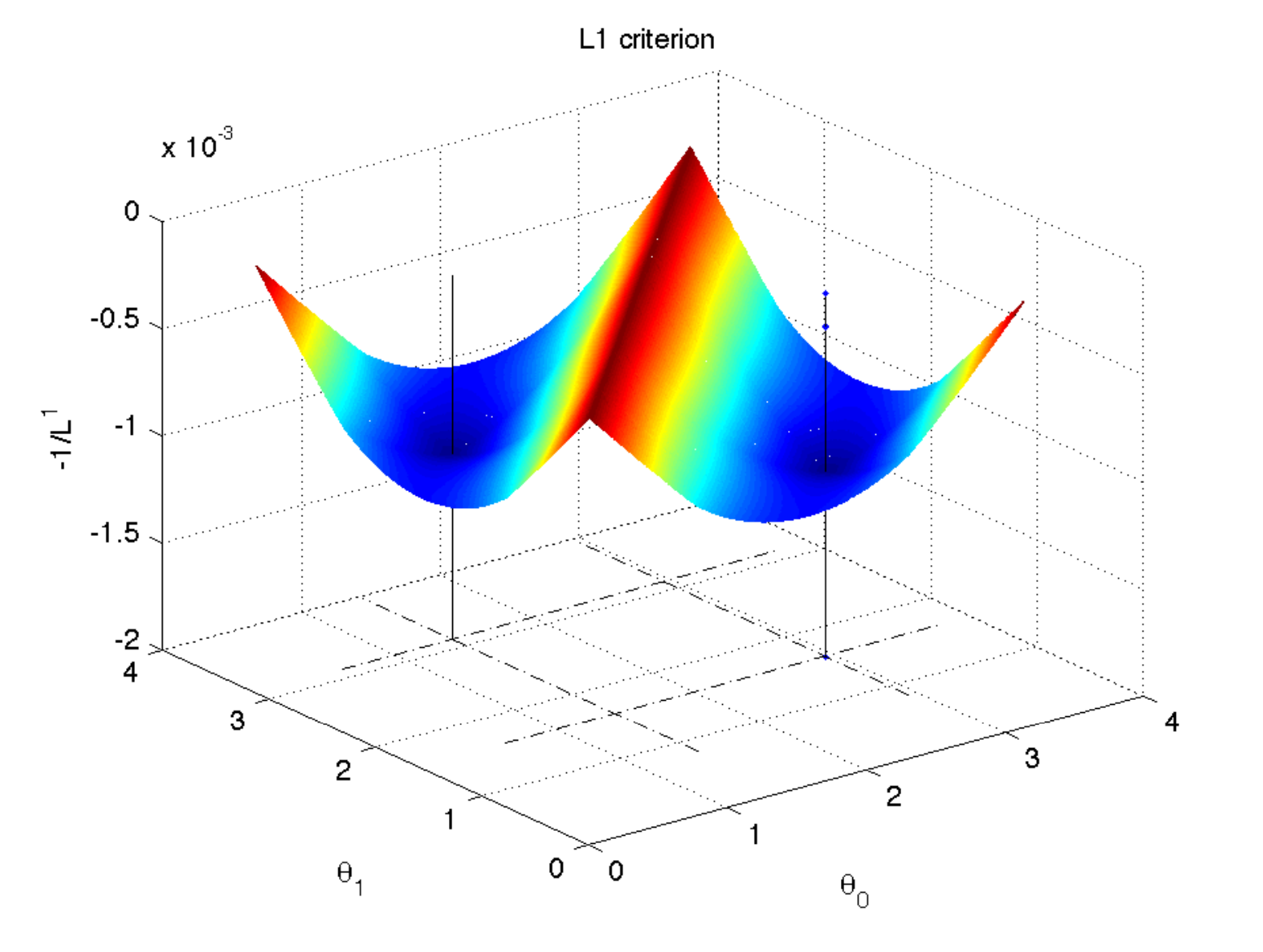}
\caption{The value of the cost $\Cost{1}(\dico| \Y)$ as a function of the angles $\theta_0$, $\theta_1$ which parameterise the dictionary $\dico = [\atom_0,\atom_1]$, $\atom_k = (\cos \theta_k, \sin \theta_k)^T \in \R^2$. Because the cost function grows to infinity when $\theta_1-\theta_0$ is close to zero, we displayed $-1/\Cost{1}(\dico|\Y)$ instead, which has the same minima.}
\label{fig:costBG}
\end{center}
\end{figure}

For the same point cloud shown on Figure~\ref{fig:cloudBG}, Figure~\ref{fig:costBG} shows the value of the $\ell_1$-cost  $\Cost{1}(\dico| \Y)$ as a function of the angles $\theta_0$, $\theta_1$ which parameterise the dictionary $\dico = [\atom_0,\atom_1]$, where $\atom_k = (\cos \theta_k, \sin \theta_k)^T \in \R^2$. One can observe that there are indeed local minima where they were expected to be located, i.e., at $(\theta_0,\theta_1) = (\theta_0^\star,\theta_1^\star)$ and $(\theta_0,\theta_1) = (\theta_1^\star,\theta_0^\star)$, which are associated to the ideal dictionary and its permuted version  (the sign ambiguity is avoided by restricting the angles to the interval $[0,\pi]$). Moreover, despite the presence of many outliers in the training data, there is no other spurious local minimum. As a result, the global minima are found where they were expected, and none is missed.

\begin{figure}[htbp]
\begin{center}
\includegraphics[width=\textwidth]{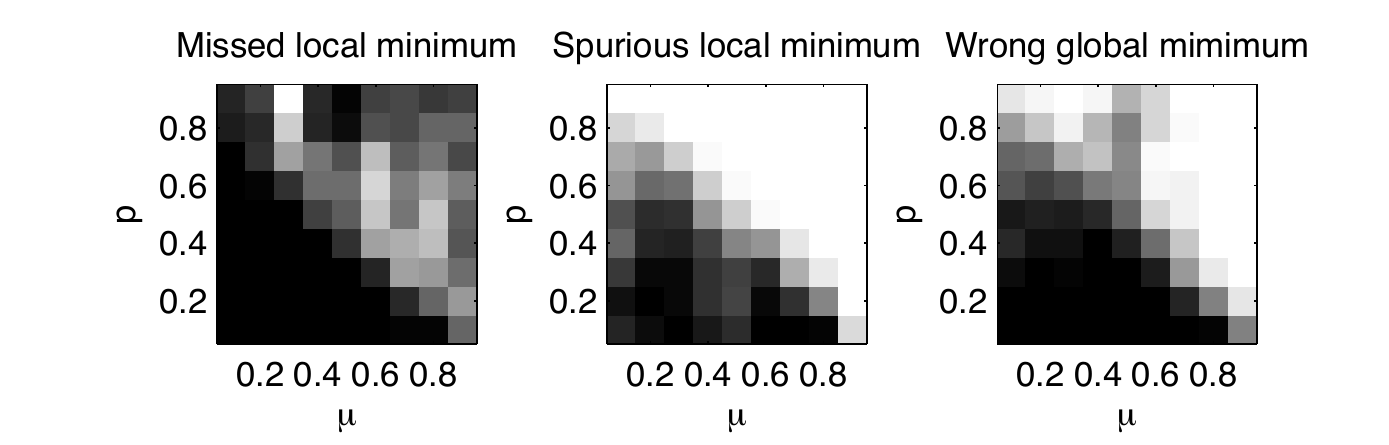}
\caption{Observed empirical phase transitions for dictionary identification by $\ell_1$ minimisation, when $\natoms = \ddim = 2$ and $\nsig$ is large. Grey level indicates observed probability of error, from black (zero) to white (one).}
\label{fig:PhaseTransitions}
\end{center}
\end{figure}

For the particular case $\natoms = \ddim = 2$, we ran a Monte-Carlo simulation where we varied the coherence $\mu$ of the dictionary and the Bernoulli-Gaussian parameter $p$ - which is associated to the typical sparsity of the generated training samples - repeating a hundred times the random draw of $\X_0$. Figure~\ref{fig:PhaseTransitions} displays the obtained results, in terms of empirical phase transitions. For small $p$ (associated to training data with many sparse samples), the black regions indicate that the probability of missing an expected  local minimum (as well as that of finding spurious one, or an erroneous global minimum) is very low, even if the coherence of the dictionary is very high. For larger values of $p$, associated to training data with more non-sparse outliers in the training set, the probability of error remains very small provided that the dictionary is sufficiently incoherent. An empirical rule of thumb seems that for small $p$, if $\mu < 1-p$ then the probability of learning errors is very small, provided that the number of training samples is sufficiently large. 

Fully characterising such phase transitions for learning over-complete dictionaries is a difficult task, for several difficulties arise at once, some due to the possible overcompleteness and non-orthogonality of the dictionary, others due to the difficulty of globally characterising the optima of a globally nonconvex problem which we know admits exponentially many solutions because of the permutation and sign indeterminacies. 
The analytic and probabilistic machinery we set up in the next sections provides tools to significantly progress towards this ambitious goal. In particular, even though the considered Bernoulli-Gaussian model may seem simplistic (it does not account for "compressible" training samples, where $x_n$ is not exactly sparse but only well approximated with few terms; neither does it account for noise $y_n = \dico_0 x_n + e_n$), we believe it is a good warm up tool to understand : a) in which conditions the $\ell_1$-criterion can be robust to non-sparse outliers; and b) whether dictionary identification is feasible using a limited number of samples. As we will see, fortunately, the answer to both questions is positive (but mathematically somewhat technical), under proper assumptions.



\section{Local Minima\label{sec:LocalIdentifiability} }

\noindent 

Instead of directly characterising the local mimina of the original problem~\eqref{eq:Learning} we consider the related problem
\begin{equation}
\label{eq:Learning1WithCoeff}
\min_{(\dico,\X) | \dico \in \manifold,  \dico \X = \Y} \|\X\|_1.\tag{P1'}
\end{equation}
It is intimately connected to the initial problem~\eqref{eq:Learning}. 
\begin{remark}\label{rem:EquivMinima}
We let the reader check the following facts.
\begin{itemize}
\item When $\dico$ is a basis ($\natoms = \ddim$), the problem~\eqref{eq:Learning1WithCoeff} is fully equivalent to the problem~\eqref{eq:Learning}, in the sense that if $\dico$ is a local (resp. global) minimum of \eqref{eq:Learning}, then the pair $(\dico,\dico^{-1}\Y)$ is a local (resp. global) minimum of ~\eqref{eq:Learning1WithCoeff}, and vice-versa.
\item When $\dico$ is overcomplete ($\natoms > \ddim$), 
\begin{itemize}
\item if $\dico$ is a local (resp. global) minimum of the original problem~\eqref{eq:Learning}, then there is a coefficient matrix $\X$ such that the pair $(\dico,\X)$ is a local (resp. global) minimum of~\eqref{eq:Learning1WithCoeff}. 
\item if $(\dico,\X)$ is a global minimum of ~\eqref{eq:Learning1WithCoeff}, then $\dico$ is a global minimum of~\eqref{eq:Learning}.  
\end{itemize}
\end{itemize}
\end{remark}

\noindent Just as in the representation problem~\eqref{eq:P1}, where the $\ell_1$-cost is not a smooth function of $x$ as soon as $x$ has at least one zero entry, the cost in Equation~\eqref{eq:Learning1WithCoeff} is not a smooth function of $(\dico,\X)$ whenever $\X$ has  at least one zero entry. Therefore, one cannot fully characterise the local minima of the cost function~\eqref{eq:Learning1WithCoeff} as a subset of the zeros of a 'gradient' of the $\ell_1$-cost function with respect to $(\dico,\X)$, for this gradient is not even well defined in a standard sense\footnote{Even the notion of G{\^a}teaux derivatives is not applicable to this cost function, which may be a reason why a standard numerical approach \cite{zipe01} is to smooth it.}. 

Here, on the opposite, we want to understand the effect of the non-smooth behaviour of the cost function, and to exploit it to characterise its local minima. For that we will develop a replacement for the 'gradient' which accounts for the fact that the $\ell_1$-cost function indeed admits one-sided directional derivatives everywhere. To keep the flow of the paper, we postpone most proofs and technical lemmata to the appendix.

\subsection{Basic Notations}
We denote by $\bad_n$ the set indexing the zero entries of the $n$-th column $x_n$ of $\X_0$, and $\bad = \{(n,k), 1 \leq n \leq \nsig, k \in \bad_n\}$ the set indexing all zero entries in $\X_0$. The notation\footnote{We will generally distinguish column vectors from row vectors using subscripts {\em vs} superscript indices.} $x^k$ is for the $k$-th row of $\X_0$, and $\bad^k$ is the set indexing the columns with a zero entry in $x^k$.\\
For any $\natoms \times \nsig$ matrix $A$ and index set $\Omega\subset \llbracket 1,\natoms \rrbracket \times \llbracket 1, \nsig \rrbracket$, the notation $A_\Omega$ will refer ubiquitously either to the vector $(A_{kn})_{(k,n) \in \Omega}$ or the $\natoms \times \nsig$ matrix which matches $A$ on $\Omega$ and is zero elsewhere. The cardinality of $\Omega$ is denoted $|\Omega|$.\\
\\

\subsection{Block Decomposition of the Considered Matrices}
In Appendix~\ref{sec:TangentSpaces} we provide a full characterisation of local minima (Lemma~\ref{LE:NSCLOCALMINIMUM}) which is sharp but somewhat abstract. To make its meaning more explicit, it is useful to consider the following block decompositions of the coefficient matrix $\X_0$ (see Figure~\ref{fig:SplitMatrix}):
\begin{figure}[ht]
\begin{center}
\includegraphics[width=\textwidth/3]{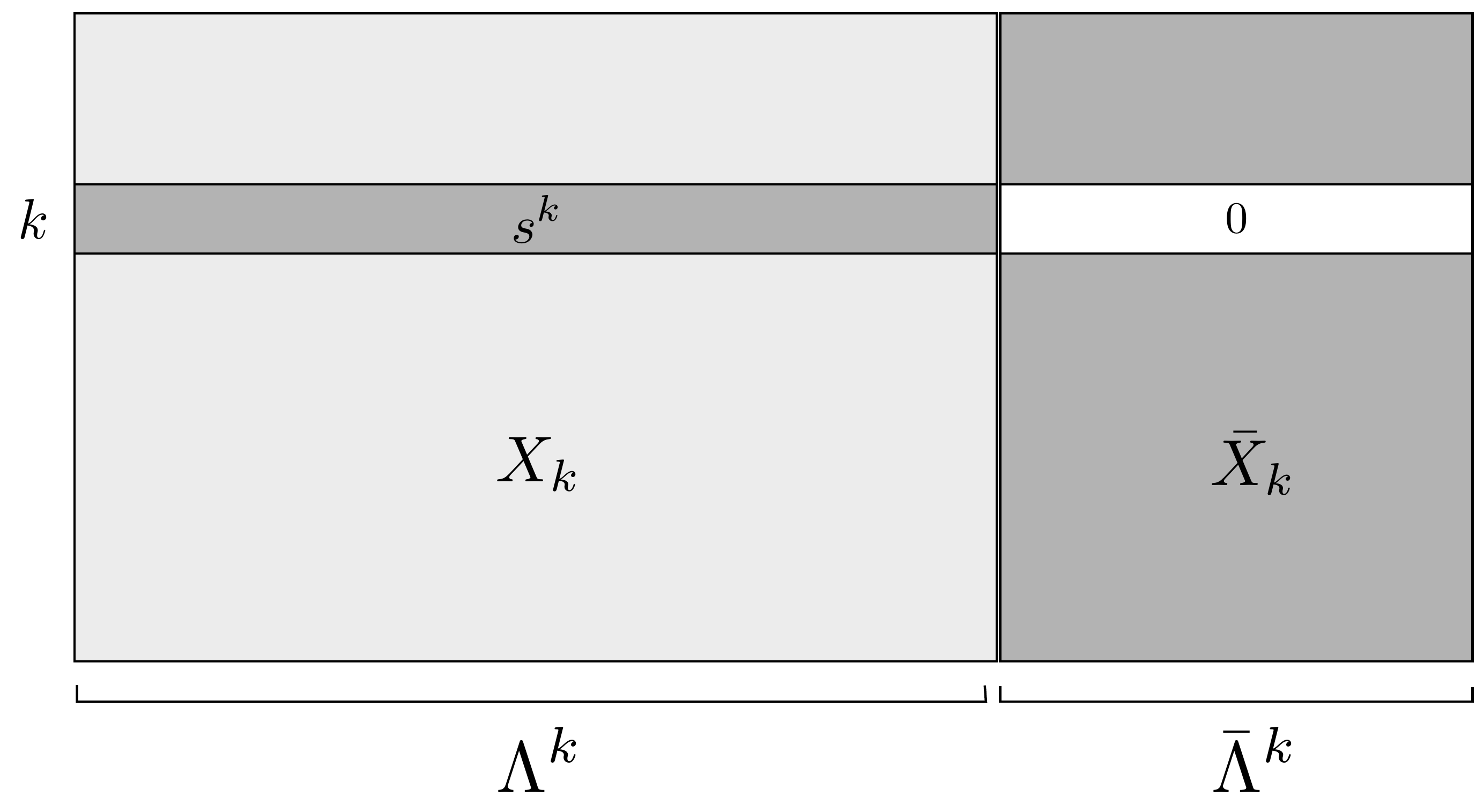}
\caption{Block decomposition of the matrix $\X_0$ with respect to a given row $x^k$. Without loss of generality, the columns of $\X_0$ have been permuted so that the first $|\good^k|$ columns hold the nonzero entries of $x^k$ while the last $|\bad^k|$ hold its zero entries.}
\label{fig:SplitMatrix}
\end{center}
\end{figure}
\begin{itemize}
\item  $x^k$  is the $k$-th row of $\X_0$;
\item $\good^k$ is the set indexing the nonzero entries of $x^k$ and $\bad^k$ the set indexing its zero entries;
\item $s^k$ is the row vector $\sign(x^k)_{\good^k}$;
\item  $\X_k$ (resp. $\bar{\X}_k$) is the matrix obtained by removing the $k$-th row of $\X_0$ and keeping only the columns indexed  by $\good^k$ (resp. $\bad^k$) .
\end{itemize}
We also define $m_k$ the $k$-th column of the 
off-diagonal part of the Gram matrix $\M = \dico_0^\star \dico_0-\Id$ and 
\begin{equation}
\label{eq:DefCoherenceRow}
\bar{m}_k 
:= 
\left(\ip{\atom_\ell}{\atom_k}
\right)_{1 \leq \ell \leq \natoms, \ell \neq k}
\end{equation}
the $k$-th column of this matrix without the zero entry corresponding to the diagonal. Finally, we consider the vectors
\begin{eqnarray}
\label{eq:DefUVect}
u_k 
&:= &
\X_k (s^k)^{\star} -  \textrm{diag}(\|x^\ell\|_1)_{1 \leq \ell \leq \natoms, \ell \neq k} \cdot  \bar{m}_k.
\end{eqnarray}

\subsection{A Necessary Condition, and a Sufficient Condition}
Equipped with these notations, we can now state the following necessary condition.
\begin{Theorem}[Necessary condition]
\label{th:NC}
Consider a complete dictionary $\dico_0 \in \manifold$, and a coefficient matrix $\X_0$ such that $\dico_0 \X_0 = \Y$. Assume that $\X_0$ is the minimum $\ell_1$ norm representation of $\Y$. With the above defined notations:
\begin{enumerate}
\item if $(\dico_0,\X_0)$ is a local minimum of~\eqref{eq:Learning1WithCoeff}; or
\item if $\dico_0$ is a global minimum of~\eqref{eq:Learning};
\end{enumerate}
then we have
\begin{equation}
\label{eq:NCRecoveryDecoupled}
\max_k \sup_{z \neq 0} \frac{|\ip{u_k}{z}|}{\|\bar{X}_k^{\star} z\|_1} \leq 1.\tag{NC}
\end{equation}
\end{Theorem} 
As a matter of fact, condition~\eqref{eq:NCRecoveryDecoupled} is almost sufficient to ensure that we have a local minimum, at least in the restricted case where $\dico_0$ is a {\em basis}, i.e., $\natoms = \ddim$. 
\begin{Theorem}[Sufficient condition, case of a basis, $\natoms=\ddim$]
\label{TH:SCRECOVERYEXPLICIT}
Consider a {\em basis} matrix $\dico_0$ with unit columns and a coefficient matrix $\X_0$ such that $\dico_0 \X_0 = \Y$. Assume that 
\begin{equation}
\label{eq:SCRecoveryDecoupled}
\max_k \sup_{z \neq 0} \frac{|\ip{u_k}{z}|}{\|\bar{X}_k^{\star} z\|_1} < 1.\tag{SC}
\end{equation}
Then $(\dico_0,\X_0)$ is a strict local minimum of~\eqref{eq:Learning1WithCoeff}.
\end{Theorem}

It remains an open question whether this type of condition is also sufficient in the case of overcomplete dictionaries. We conjecture that the answer is positive when the constant $1$ on the right hand side of~\eqref{eq:SCRecoveryDecoupled} is replaced by a sufficiently smaller value, under some additional assumptions relating the sparsity of $\X_0$ and the null space of $\dico_0$. This will be the object of further studies. For the time being, we wish to obtain a more explicit understanding of the meaning of conditions~\eqref{eq:NCRecoveryDecoupled}-\eqref{eq:SCRecoveryDecoupled}, and to characterize nontrivial collections $\X_0$ for which they are satisfied for reasonable dictionaries. In the next section we discuss the geometric interpretation of~\eqref{eq:NCRecoveryDecoupled}-\eqref{eq:SCRecoveryDecoupled}.

\section{Geometric interpretation}
\label{sec:GeometricInterpretation}
Using a duality argument (Lemma~\ref{le:Duality} in the Appendix) we first observe that for any vector $v \in \R^{\natoms-1}$, we have
\begin{equation}
\label{eq:PolytopePrimal}
\sup_{z \neq 0} \frac{|\ip{v}{z}|}{\|\bar{X}_k^{\star} z\|_1} \leq 1
\end{equation}
if, and only if, there exists a vector $d$ with $\|d\|_\infty \leq 1$ such that
$v = \bar{\X}_k d.$
In other words, condition~\eqref{eq:PolytopePrimal} holds if the vector $v \in \R^{\natoms-1}$ belongs to the convex polytope obtained by projecting the high-dimensional unit hypercube\footnote{We chose to denote the hypercube $Q$ while, technically, it depends on the considered dimension $|\bad^k|$ and will be denoted  $Q^{|\bad^k|}$ when needed.} $Q := \{ d, \|d\|_\infty \leq 1\}$ using the matrix $\bar{\X}_k$. 

The second observation is that the first summand in the definition of the vector $u_k$ (cf Eq.~\eqref{eq:DefUVect}), which is the vector 
\begin{equation}
\label{eq:DefVVect}
v_k := \X_k (s_k)^\star,
\end{equation}
is a simple weighted sum of colums of $\X_k$. Indeed, denoting $\X_k^+$ (resp. $\X_k^-$) the matrix made of the columns of $\X_k$ for which $x_n(k)$ is positive (resp. negative), the vector $v_k$ is the difference between the sum of the columns of $\X_k^+$ and the sum of those of $\X_k^-$.  

\subsection{Orthonormal Dictionaries}
\begin{figure}[htbp]
\begin{center}
\includegraphics[width=\textwidth/2]{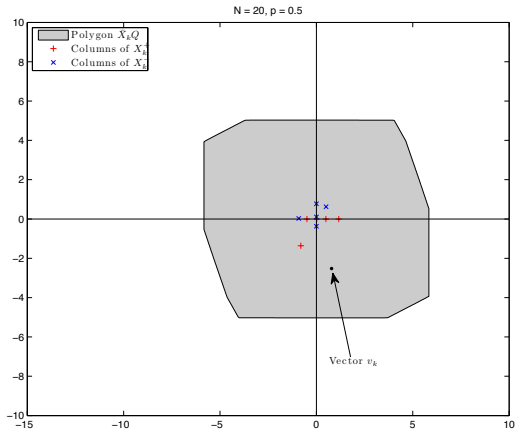}
\caption{Geometric depiction, when $\natoms = 3$, of the condition~\eqref{eq:NCRecoveryDecoupled}. The data was drawn according to the Bernoulli-Gaussian model described in Section~\ref{sec:probana}, with $p=0.5$ and $\nsig=20$. }
\label{fig:GeometricCondition1}
\end{center}
\end{figure}

\begin{figure}[htbp]
\begin{center}
\includegraphics[width=\textwidth/2]{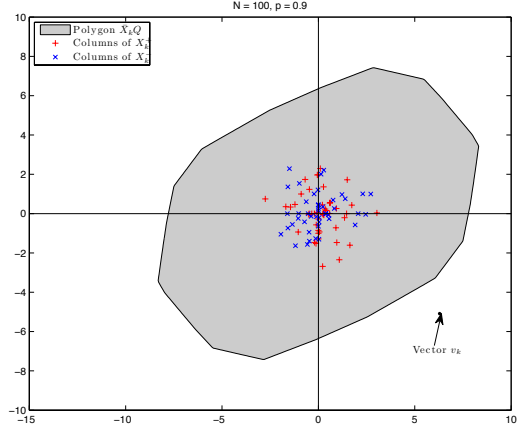}
\caption{Geometric depiction, when $\natoms = 3$, of the condition~\eqref{eq:NCRecoveryDecoupled}. The data was drawn according to the Bernoulli-Gaussian model described in Section~\ref{sec:probana}, with $p=0.9$ and $\nsig=100$. }
\label{fig:GeometricCondition2}
\end{center}
\end{figure}

Assume for a moment that the reference dictionary $\dico_0$ is an orthonormal basis. Then, we have $\M = 0$ and therefore $\bar{m}_k = 0$ and $u_k = v_k$ for all $k$. The necessary condition~\eqref{eq:NCRecoveryDecoupled} then simply reads: {\em for each $k$, the vector $v_k$ must lie within the convex polytope $\bar{X}_k Q$}. This is illustrated on Figures~\ref{fig:GeometricCondition1} and~\ref{fig:GeometricCondition2}, in dimension $\natoms = 3$, so that the vector $v_k$ as well as all the columns of $\X_k$ and $\bar{\X}_k$ live in $\R^2$. Both figures were obtained using training data  drawn according to the Bernoulli-Gaussian model described in Section~\ref{sec:probana}. 
Figure~\ref{fig:GeometricCondition1} corresponds to relatively sparse data (the parameter of the Bernoulli-Gaussian model is $p=0.5$) and we can observe that despite the relatively low number of training samples ($\nsig = 20$) the vector $v_k$ does belong to the polygon $\bar{X}_k Q$: the necessary condition~\eqref{eq:NCRecoveryDecoupled} is satisfied for the considered index $k$, and on the same data we checked that it is also satisfied for the other two indexes. Since the vectors are indeed {\em strictly} inside the considered polygons, the sufficient condition~\eqref{eq:SCRecoveryDecoupled} is also satisfied.

On the contrary, Figure~\ref{fig:GeometricCondition2} corresponds to data with many non-sparse outliers ($p=0.9$) and one can observe that despite the larger number of training samples $(\nsig = 100$), the vector $v_k$ does not belong to the polygon $\bar{X}_k Q$: the necessary condition~\eqref{eq:NCRecoveryDecoupled} is not satisfied.

\subsection{Robustness to Dictionary Coherence}

One can observe on Figure~\eqref{fig:GeometricCondition1} that the vector $v_k$ is well inside the convex polytope $\bar{\X}_k Q$. If we choose some $1 \leq \pnorm \leq \infty$, one way to quantify this fact is to say that $v_k$ has a small $\ell_\pnorm$-norm $\|v_k\|_\pnorm$ compared to the radius of the largest $\ell_\pnorm$-ball that is included in $\bar{\X}_k Q$.  
From the definition of the vector  $u_k$ (cf Eq.~\eqref{eq:DefUVect}), it follows that if the vector
\[
\textrm{diag}(\|x^\ell\|_1)_{1 \leq \ell \leq \natoms, \ell \neq k} \cdot  \bar{m}_k
\]
also has a small $\ell_\pnorm$-norm (which is the case when $\dico_0$ is not necessarily orthogonal but sufficiently "incoherent"), then $u_k$ is close to $v_k$, hence $u_k$ also lies in the polytope $\bar{\X}_k Q$. We then conclude that conditions~\eqref{eq:NCRecoveryDecoupled}-\eqref{eq:SCRecoveryDecoupled} hold true.  In other words, these conditions are robust to a certain level of dictionary coherence provided that:
\begin{enumerate}
\item each polytope $\bar{\X}_k Q$ contains a "large" $\ell_\pnorm$-ball;
\item each vector $v_k$ has "small" $\ell_\pnorm$-norm;
\item each row $x^k$ of $\X_0$ has "small" $\ell_1$-norm.
\end{enumerate}
Lemma~\ref{le:radius} in the appendix states that the radius of the largest $\ell_\pnorm$-ball included in all $\bar{\X}_k Q$ is given by
\begin{align}
\label{eq:DefAlpha}
\alpha_\pnorm(\X_0)
& :=  
\min_k 
\inf_{z \neq 0} \frac{\|\bar{\X}_k^\star z\|_1}{\|z\|_{\pnorm'}}, \\
\intertext{where  $1 \leq \pnorm' \leq \infty$ satisfies $1/\pnorm+1/\pnorm' =1$.  We also define}
\label{eq:DefBeta}
\beta_\pnorm(\X_0) 
& := \max_k \|v_k\|_\pnorm,\\
\label{eq:DefGamma}
\gamma(\X_0)
& := \max_k \|x^k\|_1.
\end{align}
We can now state the following theorem.
\begin{Theorem}
\label{TH:SCRECOVERYEXPLICITCOHERENCE}
Consider $1 \leq \pnorm \leq \infty$ and a $\natoms \times \nsig$ matrix $\X_0$.  
The conditions~\eqref{eq:NCRecoveryDecoupled}-\eqref{eq:SCRecoveryDecoupled} are satisfied provided that  the dictionary $\dico_0 \in \manifold$ is  "incoherent", in the sense that
\begin{equation}
\label{eq:DefCoher}
\mu_\pnorm(\dico_0) := \max_k \|\bar{m}_k\|_\pnorm < 
\frac{\alpha_\pnorm(\X_0)-\beta_\pnorm(\X_0)}{\gamma(\X_0)}
\end{equation}
In particular, if $\dico_0$ is an incoherent basis ($\natoms = \ddim$),  then the optimisation problem~\eqref{eq:Learning1WithCoeff} with $\Y := \dico_0 \X_0$ admits a strict local minimum at $(\dico,\X) = (\dico_0,\X_0)$.
\end{Theorem}
Compared to Theorems~\ref{th:NC} and~\ref{TH:SCRECOVERYEXPLICIT}, the above Theorem now decouples the assumptions on the coefficient matrix $\X_0$  from those on the dictionary $\dico_0$. This will considerably simplify  the analysis since we now "only" need to estimate the three quantities $\alpha_\pnorm(\X_0)$, $\beta_\pnorm(\X_0)$ and $\gamma(\X_0)$. While the last two quantities are explicit and easy to compute for a given $\X_0$, $\alpha_\pnorm(\X_0)$ is a bit more difficult to compute for a specific $\X_0$.  In Section~\ref{sec:probana}, we  show how to estimate its typical value when $\X_0$ is drawn according to a Bernoulli-Gaussian model.

\subsection{Discussion: Choice of $\pnorm$.} 
Notice that Theorem~\ref{TH:SCRECOVERYEXPLICITCOHERENCE} involves a parameter $1 \leq \pnorm \leq \infty$. One may obtain coherence conditions that may be either very restrictive on the dictionary or quite weak, depending on the choice of $\pnorm$. As we illustrate below with a few examples, the nature of the training data can have a substantial influence on the "right" choice of $\pnorm$. 
\subsubsection{Highly sparse training data}
\begin{figure}[htbp]
\begin{center}
\includegraphics[width=\textwidth/2]{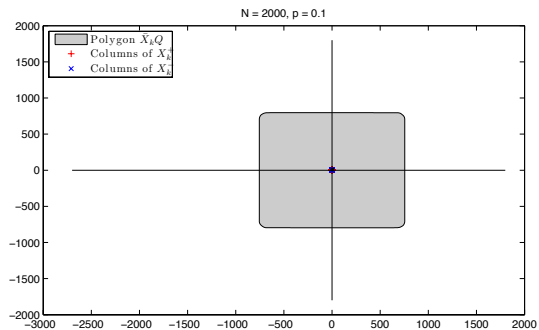}
\caption{Shape of the polytope $\bar{\X}_k Q$, $\natoms = 3$, $p=0.1$ and $\nsig=2000$. The data was drawn according to the Bernoulli-Gaussian model described in Section~\ref{sec:probana}, and is highly sparse. The shape is close to a cube.}
\label{fig:PolytopeShapeSparse}
\end{center}
\end{figure}
For a Bernoulli-Gaussian coefficient matrix $\X_0$ associated to small $p$ (highly sparse data with few non-sparse outliers), as illustrated on Figure~\ref{fig:PolytopeShapeSparse}, the polytope $\bar{\X}_k Q$ seems to be roughly shaped (when the number $\nsig$ of training samples is large) as a cube in $\R^{\natoms-1}$. Therefore, the radius of the largest included $\ell_\pnorm$-ball is almost independent of $\pnorm$, i.e., $\alpha_\pnorm(\X_0)$ is almost constant. 

Note that $\alpha_\pnorm(\X_0)$, $\beta_\pnorm(\X_0)$ and $\mu_\pnorm(\X_0)$ are always non-increasing functions of $\pnorm$. If $\alpha_\pnorm(\X_0)$ were actually constant, choosing $\pnorm = \infty$ in Eq.\eqref{eq:DefCoher} would lead to the weakest possible incoherence condition which would read in terms of the well known {\em coherence} of the dictionary
\[
\mu_\infty(\dico_0) := \max_{k \neq \ell} |\langle \atom_k,\atom_\ell\rangle| < \frac{\alpha_\infty(\X_0)-\beta_\infty(\X_0)}{\gamma(\X_0)}.
\]

\subsubsection{Almost not sparse training data}
\begin{figure}[htbp]
\begin{center}
\includegraphics[width=\textwidth/2]{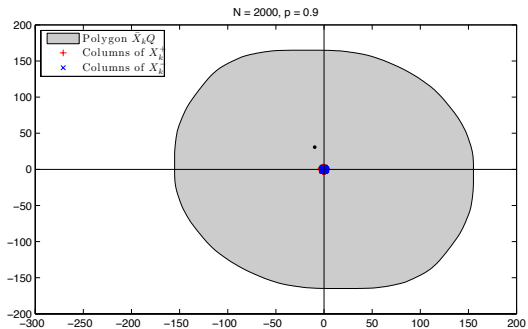}
\caption{Shape of the polytope $\bar{\X}_k Q$, $\natoms = 3$, $p=0.9$ and $\nsig=2000$. The data was drawn according to the Bernoulli-Gaussian model described in Section~\ref{sec:probana}, and is almost not sparse. The shape is close to a Euclidean ball. Note the axis coordinate which indicates that the {\em size} of the ball is somewhat smaller than in Figure~\ref{fig:PolytopeShapeSparse}, for the same number of training samples but $p = 0.1$.}
\label{fig:PolytopeShapeNonSparse}
\end{center}
\end{figure}

However, the behaviour of $\alpha_\pnorm(\X_0)$ as a function of $\X_0$ heavily depends on the nature of the training data, which determines the size and shape of the polytopes $\bar{X}_k Q$. 
Indeed, for Bernoulli-Gaussian data associated to a large $p$ (data with many non-sparse outliers), $\bar{\X}_k Q$ seems rather shaped (when $\nsig$ is large) as a Euclidean ball in $\R^{\natoms-1}$, as illustrated on Figure~\ref{fig:PolytopeShapeNonSparse}. Therefore, for such data we expect that
\[
 \alpha_\pnorm(\X_0) \approx 
 \left\{
 \begin{array}{ll}
 \alpha_2, &\pnorm \leq 2\\
 \alpha_2 \cdot (\natoms-1)^{-(1/2-1/\pnorm)}, &\pnorm \geq 2.
 \end{array}
 \right.
 \]
As a result, $\pnorm = 2$ is essentially the best choice among $1 \leq \pnorm \leq 2$, but all choices $2 \leq \pnorm \leq \infty$ remain {\em a priori} possible, depending on the behaviour of $\beta_\pnorm(\X_0)$.
\section{Probabilistic Analysis} \label{sec:probana}
In this section we will derive how many training signals are typically needed to ensure that a sufficiently incoherent basis constitutes a local minimum of the $\ell_1$-criterion, given that the coefficients of these signals are drawn from a certain probability distribution.

From a Bayesian perspective, it would seem natural to consider the Laplacian distribution: minimising the $\ell_1$-cost function corresponds to maximising the likelihood of $\dico$ under a Laplacian prior. However, when drawing coefficients from a Laplacian distribution, the probability of observing a zero entry is zero. Therefore, under the Laplacian prior, the minimum of the $\ell_1$-cost function might be close to $\dico_0$ but cannot be {\em exactly} located at $\dico_0$, no matter how many training samples are drawn. For this reason, we choose to consider coefficients drawn according to a  Bernoulli-Gaussian distribution, which ensures a nonzero probability $1-p > 0$ of observing zero entries. In a sense, the setting we consider is similar to the hypotheses of the first papers on Compressed Sensing and sparse recovery \cite{dohu01, grini03, donoho:bp}, where ill-posed linear inverse problems are solved by $\ell_1$-minimisation under an exact sparsity assumption. The difference here is that the model we consider also allows a certain proportion of non-sparse "outliers" in the training samples, as previously illustrated in Figure~\ref{fig:cloudBG}.

\subsection{The Bernoulli-Gaussian Model}\label{subsec:model}

We assume that the entries $x_{kn}$ of the $\natoms \times \nsig$ coefficient matrix $\X$ are i.i.d. with
$x_{kn}=\randsign_{kn} g_{kn}$, where the $\randsign_{kn}$ are indicator variables taking the value one with probability $p$ and zero with probability $1-p$, \ie $\randsign \sim p \delta_1 + (1-p) \delta_0$. The variables $g_{nk}$ follow a standard Gaussian distribution, \ie centered with unit variance.\\
The important role of the indicator variables is to guarantee a strictly positive probability that the entry $x_{kn}$ is exactly zero. The assumption that the $g_{nk}$ are centered Gaussians with unit variance is made mainly for simplicity reasons as it allows us to do all proofs using only elementary probability theory. However, we believe that the same results hold for many other distributions as long as they show a certain amount of concentration. 
\subsection{Asymptotic Coherence Condition}

From Theorem~\ref{TH:SCRECOVERYEXPLICITCOHERENCE} we know that we have to determine $\alpha$, $\beta$ and $\gamma$ so that with high probability
\begin{enumerate}
\item for all $k$, the image $\bar{\X}_k Q^{|\bad^k|}$ of the unit cube by the linear map $\bar{\X}_k$ contains a large $\ell_q$-ball:
\[
\alpha_q(\X_0) \geq \alpha
\]
\item for all $k$, the vector $\X_k (s^k)^{\star}$ has small $\ell_q$ norm:
\[
\beta_q(\X_0) \leq \beta,
\]
\item for all $k$, the $k$-th row $x^k$ has small $\ell_1$ norm
\[
\gamma(\X_0) \leq \gamma.
\]
\end{enumerate}
In Appendix~\ref{sec:appendixprob1}-\ref{app:probability_estimates} we derive estimates for $\alpha,\beta,\gamma$ and the associated probabilities using an $\ell_2$-ball, \ie $\pnorm=2$. Our main tools are concentration of measure results to bound the probability that a random variable deviates significantly from its expected value. We obtain probability bounds exponentially small in $\nsig$ using
\begin{eqnarray*}
\alpha & \approx & \nsig p(1-p) \textstyle \sqrt \frac 2 \pi\\
\beta & \approx & \sqrt{\nsig \natoms} p\\
\gamma & \approx & \nsig p \textstyle \sqrt \frac 2 \pi
\end{eqnarray*}
yielding, in the asymptotic regime of large $\nsig$, coherence constraints of the type
\[
\mu_2(\dico_0) < 1-p.
\]

\subsection{Non-Asymptotic Result - Required Number of Training Samples}

More specifically, we wish to quantify which number $\nsig$ of training samples guarantees, with high probability, that a basis is locally identifiable by $\ell_1$ minimisation. The following theorem, whose proof can be found in Appendix~\ref{app:proof_mth}, provides an answer to this question.
\begin{Theorem}
\label{th:maintheorem}
Let $\X$ be an $\natoms \times \nsig$ matrix drawn according to the Bernoulli-Gaussian model described in Section~\ref{subsec:model} with parameter $p < 4/5$. Assume that $\nsig > \frac{\pi\natoms}{2 (1-p)^2}$ and that $\dico_0$ is an incoherent basis such that
\begin{align}\label{eq:asympcorrbound}
\mu_2(\dico_0) < & 
1-p - \sqrt{\frac{\pi}{2}\frac{\natoms}{\nsig}}.
\end{align}
Then $\dico_0$ is locally identifiable from $\Y := \dico_0 \X$ by $\ell_1$-minimisation, except with probability at most 
\begin{align}
\label{eq:finalprobbound}
4 \natoms \exp\left( \frac{\natoms}{2} \log\left(\frac{9\natoms}{\eps^2 p}\right) - Np(1-p) \frac{\eps^2(1-2\eps)}{2}\right),
\end{align}
where $0<\eps<1/5$ is chosen as large as possible under the constraint
\begin{align}\label{eq:finalcorrbound}
\mu_2(\dico_0) \leq & 
(1-p) \cdot (1-5 \eps) \notag\\
& - \sqrt{\frac{\pi}{2}\left(\frac{\natoms}{\nsig}+\eps\right)\left(1+\frac{\eps}{p}\right)}.
\end{align}
\end{Theorem}

Note that we only require $p<4/5$ to give a simple probability bound. Similar estimates also hold for $p\geq 4/5$, see proof in Appendix~\ref{app:proof_mth}.

In the theorem above, note that we need $Np(1-p)\eps^2 > \natoms$ to have failure probability smaller than one in~\eqref{eq:finalprobbound}. The failure probability will rapidly approach zero as soon as the number of training signals $\nsig$ is larger than a constant times \[\frac{K \log K}{p(1-p)\eps^2}.\] Considering that, in order not to have a trivial sparse solution, where the columns of $\dico$ are scaled versions of the training samples $y_n$, we need at least $K+1$ training samples, this is not a large requirement. 

\noindent {\bf Example:} consider $\dico_0$ a basis of $\R^\natoms$ made of $1 \leq \ell \leq \natoms/2$ (resp. $\natoms-\ell$) vectors from an orthonormal basis $\dico_1$ (resp. $\dico_2$) where $\dico_2$ is maximally incoherent with $\dico_1$ \cite{dohu01,grini03}. It is easy to check that $\mu_2(\dico_0) = 1-\ell/\natoms < 1$, hence $\dico_0$ is, with high probability, a local minimum of the $\ell_1$-criterion with $\Y = \dico_0 \X_0$ when $\X_0$ is drawn according to the Bernoulli Gaussian model with $p < \ell/\natoms < 1/2$.

\section{Discussion} \label{sec:discussion}
We have developed necessary and sufficient algebraic conditions on a dictionary coefficient pair to constitute a local minimum of the $\ell_1$-dictionary learning criterion. In case the dictionary is an incoherent basis we have shown that for coefficient matrices generated from a random sparse model the resulting basis coefficient pair suffices these conditions with high probability as long as the number of training signals grows like $\ddim \log \ddim$. These are exciting new results but since dictionary learning is a relatively young field they lead to more open questions. 

For the special case when the dictionary is assumed to be a basis a helpful result for practical purposes would be to prove that under the random model there exists only one local minimum which then has to be the global one, and could be found with simple descent algorithms. Numerical experiments in two dimensions support this hypothesis, as shown in Figure~\ref{fig:costBG} where the only two local minima are at the original dictionary $\dico_0$ and at the dictionary corresponding to $\dico_0$ with permuted columns.


It would be also desirable to show the converse direction, \ie if the coherence of the basis is too high and the training signals are generated by the same random sparse model, the basis coefficient pair will not be a local minimum. Again, this is empiricaly the case as shown in Figure~\ref{fig:PhaseTransitions}. To answer this question from a theoretical perspective, it will first be necessary to investigate for which $q$ the $\ell_q$-ball most resembles the image of the unit cube under $\bar{\X}_k$. In the proof here we used $q=2$ but there are some indications that $q=\infty$ is the more appropriate choice, which could also lead to a sharper version of the current result. Ideally we could then show that, as soon as a basis has coherence $\max_k \|m_k\|_q$ higher than $(1-p)$, it is extremely unlikely to be a local minimum. 

Finally much harder research will have to be invested to extend the current results to the overcomplete and the noisy case. In the overcomplete case, the null space has to be taken into account, which prevents a straightforward generalisation from the intrinsic necessary and sufficient conditions of Lemma~\ref{LE:NSCLOCALMINIMUM} to explicit sufficient conditions as in Theorem~\ref{TH:SCRECOVERYEXPLICIT}. In the noisy case, even the formulation of the problem has to be changed as we cannot expect the best dictionary for the noise contaminated training data to be exactly the same as the original dictionary but only close to it.

\appendices

\section{Notations \label{sec:notations}} 
\noindent To state the main lemmata  we need to introduce the following notation conventions.\\
\\
\noindent
{\bf Froebenius norm and inner product.}\\
For any matrix, $A^\star$ denotes the transpose of $A$. We let \(
\ip{A}{B}_F = \trace(A^{\star}B) 
\)
denote the natural inner product between matrices, which is associated to the Froebenius norm $\|A\|_F^2 = \ip{A}{A}_F$, and $\sign(A)$ is the $\sign$ operator applied componentwise to the matrix $A$ (by convention $\sign(0) := 0$). All proofs will rely extensively on the fact that 
\begin{equation}
\label{eq:PropFrobenius}
\ip{AB}{C}_F = \trace(B^{\star}A^{\star}C) = \trace(A^{\star}CB^{\star}) = \ip{A}{CB^{\star}}_F
\end{equation}
and similar relations such as
\begin{equation}
\label{eq:PropDiagAdjoint}
\ip{\diag(A)}{B}_F = \ip{A}{\diag(B)}_F.
\end{equation}
\\
\noindent
{\bf Zero-diagonal \& diagonal decomposition.}\\
We will use  the following simple lemma.
\begin{lemma}
\label{le:zerodiag}
Consider $\mathbf{A}$,$\mathbf{B}$  two matrices and let $\mathbf{A} = \Zdiag_1 + \Diag_1$, $\mathbf{B} =  \Zdiag_2+\Diag_2$ be their unique decomposition into a sum of a zero-diagonal and a diagonal matrix. Then 
\[
\diag(\mathbf{A}\mathbf{B}) = \Diag_1 \Diag_2 + \diag(\Zdiag_1 \Zdiag_2).
\]
\end{lemma}
\begin{Proof}
The product of a zero-diagonal matrix with a diagonal matrix is zero-diagonal, hence $\Zdiag_1 \Diag_2$ and $\Diag_1 \Zdiag_2$ are zero-diagonal and
\begin{eqnarray*}
\diag(\mathbf{A}\mathbf{B}) 
&=& \diag\left((\Zdiag_1+\Diag_1)(\Zdiag_2+\Diag_2)\right) \\
&=& \diag\left(\Zdiag_1\Zdiag_2 +\Diag_1\Zdiag_2 + \Zdiag_1 \Diag_2 + \Diag_1 \Diag_2 \right)\\
&=& \diag(\Zdiag_1\Zdiag_2)  + \Diag_1 \Diag_2.
\end{eqnarray*}
\end{Proof}

For any dictionary $\dico_0 \in \manifold$, we will consider in particular the decomposition of the Gram matrix $\dico_0^{\star} \dico_0$ into a diagonal part and a zero-diagonal part:
\begin{eqnarray}
\label{eq:GramDiag}
\Diag_0 & := & \diag(\dico_0^{\star} \dico_0) = \diag(\|\atom_k\|_2^2) = \Id,\\
\label{eq:GramOffdiag}
\M       & := & \dico_0^{\star}\dico_0 - \Id.
\end{eqnarray}
\\

\noindent
{\bf Null space}\\
We denote by $\Null(\dico)$ the null space of the dictionary $\dico$, \ie the linear subspace made up of all column vectors $v \in \R^\natoms$ such that $\dico v = 0$. By abuse of notation, we will also denote $\Null(\dico)$ the linear space of $\natoms \times \nsig$ matrices $\V$ such that $\dico \V = 0$.\\
\\
\noindent
{\bf $\mathbf{\eps}$-cover}\\
A finite $\eps$-cover of the unit $\ell^\pnorm$-sphere in $\R^n$ is a finite set $\epscover$ of points with unit $\ell^\pnorm$-norm such that for all points in the sphere, \ie $\|x\|_\pnorm=1$, we have
\begin{align}
\min_{x_i \in\epscover} \|x-x_i\|_\pnorm < \eps.\notag
\end{align} 
From Lemma~4.10 in \cite{Pisier:1999aa} we know that for $\eps \in (0,1)$ there always exists an $\eps$-cover $\epscover$ with cardinality $|\epscover|<(3/\eps)^n$.

\section{Tangent spaces and local minima}
\label{sec:TangentSpaces}

To characterise whether $(\dico_0,\X_0)$ is a local minimum of~\eqref{eq:Learning1WithCoeff}, we will use the notion of  the tangent space $\tangentspace{\mathcal{M}(\Y)}{(\dico_0,\X_0)}$ to the constraint manifold
\begin{equation}
\label{eq:DicoCoeffManifold}
\mathcal{M}(\Y) := \{(\dico,\X), \dico \in \manifold, \dico \X = \Y\}
\end{equation}
at the point $(\dico_0,\X_0)$. We characterise this tangent space before providing the characterisation of the local minima.

\subsection{The Tangent Space $\tangentspace{\mathcal{M}(\Y)}{(\dico_0,\X_0)}$\label{sec:TangentSpace}}
\noindent The tangent space $\tangentspace{\mathcal{M}(\Y)}{(\dico_0,\X_0)}$ to the constraint manifold $\mathcal{M}(\Y)$ at the point $(\dico_0,\X_0)$ is the collection of the derivatives $(\dico',\X') :=  (\dico'(0),\X'(0))$ of all smooth functions $\epsilon \mapsto (\dico(\epsilon),\X(\epsilon))$ which satisfy $\forall \epsilon, (\dico(\epsilon),\X(\epsilon)) \in \mathcal{M}(\Y)$ and $(\dico(0),\X(0)) = (\dico_0,\X_0)$. \\
Below we characterise the tangent spaces $\tangentspace{\manifold}{\dico_0}$ and $\tangentspace{\mathcal{M}(\Y)}{(\dico_0,\X_0)}$. The characterisations use the decomposition $\dico_0^{\star}\dico_0 = \Id + \M$ introduced in Equations~\eqref{eq:GramDiag}-\eqref{eq:GramOffdiag}, through the notion of {\em admissible} matrices: a square $\natoms \times \natoms$ matrix $C$ is said to be admissible if $\dico' := \dico_0 \cdot C \in \tangentspace{\manifold}{\dico_0}$. 
\begin{lemma}
 \label{le:TangentSpace}
 Let $\dico_0 \in \manifold$ be a complete dictionary. 
  \begin{itemize}
\item Any matrix $\dico ' \in \tangentspace{\manifold}{\dico_0}$ can be written as $\dico' = \dico_0 \cdot C$ for some admissible $C$. 
\item The matrix $C$ is admissible if, and only if there exists a zero-diagonal matrix $\Zdiag$ 
such that 
\begin{equation}
\label{eq:AdmissibleMatrix}
C = \Zdiag - \diag(\M \Zdiag) 
\end{equation}
\end{itemize}
\end{lemma}

\begin{Proof}
The first claim is a trivial consequence of the completeness of $\dico_0$, which shows that any matrix can be written as $\dico_0 \cdot C$, and the definition of an admissible matrix.\\
The constraint in~\eqref{eq:DicoConstraintUnit} 
can be rewritten as 
$
\diag(\dico^{\star} \dico) = \Id.
$
Taking the derivative, it follows that $\dico' \in \tangentspace{\manifold
}{\dico_0}$ if, and only if, 
$
\diag(\dico_0^{\star}\dico') = 0.
$
Writing $\dico' = \dico_0 \cdot C$ and decomposing  $C = \Zdiag + \Diag$ into a zero-diagonal and a diagonal matrix, we obtain from Lemma~\ref{le:zerodiag}
\begin{eqnarray*}
\diag(\dico_0^{\star}\dico') 
&=& \diag(\dico_0^{\star} \dico_0\cdot C) = \diag\left( (\M+\Id)(\Zdiag + \Diag)\right)\\
&=& \Diag + \diag(\M\Zdiag).
\end{eqnarray*}
Hence $\dico_0 \cdot C \in \tangentspace{\manifold
}{\dico_0}$ if, and only if, $\Diag = - \diag(\M\Zdiag)$, \ie  if $C = \Zdiag - \diag(\M\Zdiag)$. 

\end{Proof}

\begin{lemma}
\label{le:TangentSpaceFull}
The pair $(\dico',\X')$ is in the tangent space $\tangentspace{\mathcal{M}(\Y)}{(\dico_0,\X_0)}$ if, and only if, there exists an arbitrary admissible matrix $C$ and an arbitrary element $\V$ of $\Null(\dico_0)$ such that
\begin{eqnarray}
\label{eq:TangentSpaceFull1}
\dico' &=& \dico_0 \cdot C\\
\label{eq:TangentSpaceFull2}
\X' &=& -C \X_0 + \V. 
\end{eqnarray}
\end{lemma}

\begin{Proof}
Given the nature of the constraint manifold $\mathcal{M}(\Y)$, its tangent space at $(\dico_0,\X_0)$ is made up of all the pairs $(\dico',\X')$ such that $\dico' \in \tangentspace{\manifold}{\dico_0}$ and
$\dico' \X_0 + \dico_0 \X' = 0$, meaning $\dico' = \dico_0 \cdot C$ with some admissible $C$, and
 $\dico_0 (C \X_0 + \X') = 0$. The latter is equivalent to  $C \X_0 + \X' \in \Null(\dico_0)$. 
\end{Proof}

\subsection{Characterisation of Local Minima\label{sec:LocalMinimaAndGradient}}

\begin{lemma}
\label{LE:NSCLOCALMINIMUM}
Consider a complete dictionary $\dico_0 \in \manifold$, and a coefficient matrix $\X_0$ such that $\dico_0 \X_0 = \Y$. Define the $\natoms \times \natoms$ matrix
\begin{equation}
\label{eq:DefUMatrix}
\mathbf{U} := \sign(\X_0)\X_0^{\star} - \M^{\star} 
\diag(\|x^k\|_1).
\end{equation}
\begin{enumerate}
\item \label{it:sufficientlocmin} If for every zero-diagonal $\Zdiag$ and $\V \in \Null(\dico_0)$ such that $\Zdiag \X_0 + \V \neq 0$ we have
\begin{equation}
\label{eq:NSCRecovery}
\left|\ip{\Zdiag}{\mathbf{U}}_F 
+
\ip{\V}{\sign(\X_0)}_F\right| < \|(\Zdiag \X_0 + \V)_\bad\|_1,
\end{equation}
then $(\dico_0,\X_0)$ is a strict local minimum of~\eqref{eq:Learning1WithCoeff}. 
\item \label{it:necessarylocmin} If the reversed strict inequality holds in~\eqref{eq:NSCRecovery}  for some 
 zero-diagonal $\Zdiag$ and some $\V \in \Null(\dico_0)$ such that $\Zdiag \X_0 + \V \neq 0$, then $(\dico_0,\X_0)$ is {\em not} a local minimum of~\eqref{eq:Learning1WithCoeff}.
\end{enumerate}
\end{lemma}

\begin{Proof}
Denote $a(\epsilon) \doteq b(\epsilon)$ when $\lim_{\epsilon \to 0}\|a(\epsilon)-b(\epsilon)\|/|\epsilon| = 0$. Consider any smooth function $\epsilon \mapsto (\dico(\epsilon),\X(\epsilon)) \in \mathcal{M}(\Y)$. By definition we have 
\(
\X(\epsilon)  \doteq \X_0+\epsilon \X',
\)
and for small $\epsilon$, the sign of $\X(\epsilon)$ matches that of $\X_0 = \X(0)$ on the support $\good$ of $\X_0$, hence we may write 
\begin{eqnarray*}
\|\X\|_1 
&=& \ip{\X}{\sign(\X)}_F\\
&=& \|(\X-\X_0)_{\bad}\|_1+ \ip{\X}{\sign(\X_0)}_F\\
& = & \|(\X-\X_0)_{\bad}\|_1\\
&& + \ip{\X-\X_0}{\sign(\X_0)}_F + \|\X_0\|_1,\\
\|\X\|_1-\|\X_0\|_1 &=& \|(\X-\X_0)_\bad\|_1 + \ip{\X-\X_0}{\sign(\X_0)}_F\\
&\doteq& 
|\epsilon| \cdot \|(\X')_\bad\|_1 + \epsilon \ip{\X'}{\sign(\X_0)}_F.
\end{eqnarray*}
As a result, the one-sided derivatives of the $\ell_1$-criterion in the tangent direction $(\dico',\X')$ are
\begin{eqnarray*}
\label{eq:RightSidedDirectionalDerivative}
\nabla_{\dico',\X'}^+ \|\X\|_1 
&:=& 
\lim_{\epsilon \to 0, \epsilon > 0} \frac{\|\X(\epsilon)\|_1-\|\X_0\|_1}{\epsilon} \\
&=&  
+\|(\X')_\bad\|_1 + \ip{\X'}{\sign(\X_0)}_F\nonumber\\
\label{eq:LeftSidedDirectionalDerivative}
\nabla_{\dico',\X'}^- \|\X\|_1 
&:=& 
\lim_{\epsilon \to 0, \epsilon < 0} \frac{\|\X(\epsilon)\|_1-\|\X_0\|_1}{\epsilon} \\
&=&  
-\|(\X')_\bad\|_1 + \ip{\X'}{\sign(\X_0)}_F,\nonumber
\end{eqnarray*}
and the $\ell_1$-criterion admits a local minimum at $(\dico_0,\X_0)$ if for all $(\dico',\X')$ in the tangent space $\tangentspace{\mathcal{M}(\Y)}{(\dico_0,\X_0)}$ with $\X' \neq 0$ we have 
\begin{equation*}
\label{eq:GenericLocalMinConditionL1}
|\ip{\X'}{\sign(\X_0)}_F| <  \|(\X')_\bad\|_1.
\end{equation*}
Vice-versa,  the $\ell_1$-criterion does not admit a local minimum at $(\dico_0,\X_0)$ if there exists some $(\dico',\X')$ in the tangent space $\tangentspace{\mathcal{M}(\Y)}{(\dico_0,\X_0)}$ yielding the reversed strict inequality.\\
Using Lemma~\ref{le:TangentSpaceFull} we get that the $\ell_1$-criterion admits a local minimum at $(\dico_0,\X_0)$ if for all admissible $C$  and all $\V \in \Null(\dico_0)$ such that $\V \neq C\X_0$ we have
\begin{equation}
\label{eq:LocalMinConditionL1}
|\ip{C\X_0+\V}{\sign(\X_0)}_F| <  \|(C\X_0+\V)_\bad\|_1.
\end{equation}
The rest of the proof consists in rewriting~\eqref{eq:LocalMinConditionL1} using Lemma~\ref{le:TangentSpace} and the properties~\eqref{eq:PropFrobenius} and~\eqref{eq:PropDiagAdjoint}.\\
First, using~\eqref{eq:PropFrobenius}, the inequality in~\eqref{eq:LocalMinConditionL1}  is equivalent to 
\[
\left|\ip{C}{\sign(\X_0)\X_0^{\star}}_F + \ip{\V}{\sign(\X_0)}_F\right|
<
 \| (C \X_0+\V)_\bad \|_1.
\]
Second, by Lemma~\ref{le:TangentSpace}, the admissible matrices are exactly the matrices 
\(
C = \Zdiag - \diag(\M \Zdiag)
,
\)
with $\Zdiag$ an arbitrary zero-diagonal matrix
. Since $(\Diag \cdot \X_0)_\bad = 0$ for any diagonal matrix $\Diag$, we get $(C\X_0)_\bad = (\Zdiag \X_0)_\bad$ for any admissible matrix. The inequality is therefore equivalent to
\begin{eqnarray}
& & |\ip{\Zdiag - \diag(\M \Zdiag)}{\sign(\X_0)\X_0^{\star}}_F
+ 
\ip{\V}{\sign(\X_0)}_F|\notag\\
\label{eq:LocalMinConditionL12}
&& < \|(\Zdiag \X_0 + \V)_\bad\|_1,
\end{eqnarray}
with arbitrary zero-diagonal $\Zdiag$ and $\V \in \Null(\dico_0)$.\\
Third, since $\diag(\sign(\X_0)\X_0^{\star}) = \diag(\|x^k\|_1)$, we observe using~\eqref{eq:PropFrobenius} and~\eqref{eq:PropDiagAdjoint} that
\begin{align}
\ip{\diag(\M\Zdiag)}{\sign(\X_0)\X_0^{\star}}_F
&=
\ip{\M\Zdiag}{\diag(\sign(\X_0)\X_0^{\star})}_F\\
&=
\ip{\Zdiag}{\M^{\star} \diag(\|x^k\|_1)}_F.\notag
\end{align}
Hence the inequality in~\eqref{eq:LocalMinConditionL12} is equivalent to 
\begin{align*}
\big|\ip{\Zdiag}{\sign(\X_0)\X_0^{\star} - \M^{\star} \diag(\|x^k\|_1)}_F 
&+
\ip{\V}{\sign(\X_0)}_F\big|\\
& < \|(\Zdiag \X_0 + \V)_\bad\|_1.
\end{align*}
\end{Proof}
\subsection{Proof of Theorems~\ref{th:NC} and~\ref{TH:SCRECOVERYEXPLICIT}}

\begin{lemma}
\label{LE:OBLIQUELEARNING}
Using the notations of Section~\ref{sec:LocalIdentifiability} we have
\begin{equation}
\label{eq:NSCRecoveryBasisDecoupled}
\sup_{\Zdiag \neq 0} \frac{\big|\ip{\Zdiag}{\mathbf{U}}\big|}{\big\|(\Zdiag \X_0)_\bad\big\|_1}
=
\max_k \sup_{z \in \R^{\natoms-1}\backslash\{0\}},\  \frac{|\ip{u_k}{z}|}{\|\bar{X}_k^{\star} z\|_1}.
\end{equation}
\end{lemma}

\begin{Proof}
Denote $z^k$ the $k$-th row of the zero diagonal matrix $\Zdiag$: it is a row vector in $\R^\natoms$ with a zero entry at the $k$-th coordinate, and we denote $\bar{z}^k$ the row vector in $\R^{\natoms-1}$ obtained by removing this zero entry.  Observe that the $k$-th row of $\Zdiag \X_0$ is $z^k \X_0 = \bar{z}^k \X_0^k$ where $\X_0^k$ is $\X_0$ with the $k$-th row removed. As a consequence the denominator in Eq.~\eqref{eq:NSCRecoveryBasisDecoupled} is decomposed into the sum
\begin{align}
\|(\Zdiag \X_0)_\bad\|_1 
& = 
\sum_k  \|(z^k \X_0)_{\bad^k}\|_1
=
\sum_k  \|(\bar{z}^k \X_0^k)_{\bad^k}\|_1\notag\\
& =
\sum_k  \|\bar{z}^k (\X_0^k)_{\bad^k}\|_1
=
\sum_k  \|\bar{z}^k \bar{\X}_k\|_1.
\end{align}
Now we decompose the numerator into a similar sum.  First, we observe that
\begin{align*}
 \ip{\Zdiag}{\M^{\star} \diag(\|x^k\|_1)}_F
&=
 \sum_k \ip{z^k}{ m_k^\star \, \textrm{diag}(\|x^\ell\|_1)_{1 \leq \ell \leq \natoms}   }\\
& = 
\sum_k \cdot \ip{\bar{z}^k}{  \bar{m}_k^\star \, \textrm{diag}(\|x^\ell\|_1)_{1 \leq \ell \leq \natoms, \ell \neq k} },\\
\ip{\Zdiag}{\sign(\X_0)\X_0^{\star}}_F
&=
\ip{\Zdiag\X_0}{\sign(\X_0)}_F \\ 
&=
 \sum_k \ip{z^k \X_0}{\sign(x^k)}\\
& =
 \sum_k \ip{\bar{z}^k \X_0^k}{\sign(x^k)}.
\end{align*}
Then, by matching  column permutations  of $\X_0^k$ and $\sign(x^k)$ we get
\begin{align*}
\ip{\bar{z}^k \X_0^k}{\sign(x^k)} 
= 
\ip{\bar{z}^k [\X_k;\bar{\X}_k]}{[s^k;0]} 
&= 
\ip{\bar{z}^k \X_k}{s^k}\\
&=
\ip{\bar{z}^k }{s^k\X_k^{\star}},
\end{align*}
and conclude that the numerator is 
\begin{equation}
\big|\ip{\Zdiag}{\mathbf{U}}\big| = \big|\sum_k  \ip{\bar{z}^k }{u_k^\star}\big|.
\end{equation}
The conclusion is then straightforward.
\end{Proof}
\begin{Proof}[Proof of Theorem~\ref{th:NC}]
Using Lemma~\ref{LE:NSCLOCALMINIMUM} and Remark~\ref{rem:EquivMinima} we know that  
if $\dico_0$ is a local minimum of~\eqref{eq:Learning1WithCoeff} or a global minimum of~\ref{eq:Learning1}, then for any zero-diagonal matrix $\Zdiag$ and any $\V \in \Null(\dico_0)$ such that $\Zdiag \X_0 + \V \neq 0$ we have 
$\big|\ip{\Zdiag}{\mathbf{U}} + \ip{\V}{\sign(\X_0)}\big| \leq \big\|(\Zdiag \X_0+\V)_\bad\big\|_1$. In particular, for any $\Zdiag \neq 0$ and $\V = 0$, we have
$\big|\ip{\Zdiag}{\mathbf{U}}\big| \leq \big\|(\Zdiag \X_0)_\bad\big\|_1$. 
We conclude using Lemma~\ref{LE:OBLIQUELEARNING}.
\end{Proof}

\begin{Proof}[Proof of Theorem~\ref{TH:SCRECOVERYEXPLICIT}]
When $\dico_0$ is a basis, the null space is $\Null(\dico_0) = \{0\}$, and Condition~\eqref{eq:NSCRecovery} is satisfied for all nonzero zero-diagonal matrices $\Zdiag$ and $\V \in \Null(\dico_0)$ such that $\Zdiag \X_0 + \V \neq 0$ if, and only if, for all nonzero zero-diagonal matrix  $\Zdiag$ we have
$|\ip{\Zdiag}{\mathbf{U}}_F| < \|(\Zdiag \X_0)_\bad\|_1.$
Again, we conclude thanks to Lemma~\ref{LE:OBLIQUELEARNING}.
\end{Proof}

\subsection{Duality Analysis}
\noindent The next lemma exploits duality to understand the geometric meaning of conditions in~\eqref{eq:NCRecoveryDecoupled}-\eqref{eq:SCRecoveryDecoupled}. The following Lemma is used with the matrix $A = \bar{\X}_k$ to obtain the equivalent characterization of \eqref{eq:PolytopePrimal} used in Section~\ref{sec:GeometricInterpretation}.
\begin{lemma}
\label{le:Duality}
Let $A$ be an $n \times M$ matrix with rank $n$. For any vector $v$ define
\begin{eqnarray}
\label{eq:DefDualNorm}
\|v\|_A 
&:= &
\sup_{z \neq 0} \frac{\ip{v}{z}}{\|A^\star z\|_1}.
\end{eqnarray}
We have the equivalent characterisation
\begin{equation}
\label{eq:DualityProperty}
\|v\|_A = \min \|d\|_\infty,\ \mbox{under the constraint}\ A d = v.
\end{equation}
\end{lemma}

\begin{Proof}
We will just prove that 
\[
\|v\|_A \leq \min \|d\|_\infty,\ \mbox{under the constraint}\ A d = v.
\]
The reversed inequality is more technical but only requires casting both norm characterisations~\eqref{eq:DefDualNorm}-\eqref{eq:DualityProperty} to a pair of linear programs in primal and dual form, and using the strong duality theorem to show that both programs, which are bounded and feasible, have the same value of the optimum.
To check the easy inequality, take any $d$ such that $A d = u$ . Since $A$ has rank $n$,  we have $\|A^\star z\|_1 \neq 0$ whenever $z \neq 0$. Thus, for any $z \neq 0$ we have
$
\ip{v}{z}
= 
\ip{A d}{z}
= 
\ip{d}{A^\star z}
\leq 
\|d\|_\infty \cdot \|A^\star z\|_1,
$
hence $\|v\|_A \leq \|d\|_\infty$. 
\end{Proof}

\begin{lemma}
\label{le:radius}
Consider $A$ an $n \times M$ matrix and $1 \leq \pnorm,\pnorm' \leq \infty$ with $1/\pnorm+1/\pnorm' =1$. The radius of the largest $\ell_\pnorm$ ball included in $A Q^M$ is
\begin{equation}
\radius_\pnorm(A) := \inf_{z \neq 0} \frac{\|A^\star z\|_1}{\|z\|_{\pnorm'}}.
\end{equation}
\end{lemma}
\begin{Proof}
If $A$ is not of rank $n$ we let the reader check that $\radius_\pnorm(A) = 0$ is also the radius of the largest ball included in $A Q^M$. Otherwise, from Lemma~\ref{le:Duality} we know that $v  \in A Q^M$ if and only if $\sup_{z \neq 0} \frac{|\ip{v}{z}|}{\|A^\star z\|_1} \leq 1$. The inclusion of an $\ell_\pnorm$ ball of radius $\alpha$ in $A Q^N$ is therefore equivalent to
\begin{align*}
\sup_{\|v\|_\pnorm \leq \alpha} \sup_{z \neq 0}  \frac{|\ip{v}{z}|}{\|A^\star z\|_1} 
&\leq 1.\\
\intertext{Conclude by rewriting the left hand side: }
\alpha \sup_{\|v'\|_\pnorm \leq 1, z \neq 0}  \frac{|\ip{v'}{z}|}{\|A^\star z\|_1}
& = 
\alpha \sup_{z \neq 0}  \frac{\|z\|_{\pnorm'}}{\|A^\star z\|_1}.
\end{align*}
\end{Proof}
\subsection{Proof of Theorem~\ref{TH:SCRECOVERYEXPLICITCOHERENCE}}
\noindent Using the definition of $u_k$, $\beta_\pnorm(\X_0)$, $\gamma(\X_0)$ and $\mu_\pnorm(\dico_0)$ (cf Eqs.~\eqref{eq:DefUVect},~\eqref{eq:DefBeta},~\eqref{eq:DefGamma} and ~\eqref{eq:DefCoher}) and the assumption on $\mu_\pnorm(\dico_0)$ (Eq.~\eqref{eq:DefCoher}) we have
for all $k$ 
\begin{align*}
\|u_k\|_\pnorm
& \leq
\|v_k\|_\pnorm + \gamma(\X_0) \cdot \mu_\pnorm(\dico_0)\\
& \leq \beta_\pnorm(\X_0) + \gamma(\X_0) \cdot \mu_\pnorm(\dico_0) < \alpha_\pnorm(\X_0).
\end{align*}
Hence, by definition of $\alpha_\pnorm(\X_0)$ the vector $u_k$ belongs to $\bar{\X}_k Q$ for all $k$,
and we conclude using Lemma~\ref{le:Duality} that the condition~\eqref{eq:SCRecoveryDecoupled} is satisfied. In particular, if $\dico_0$ is a basis then we conclude using Theorem~\ref{TH:SCRECOVERYEXPLICIT} that $(\dico_0,\X_0)$ is a local minimum of~\eqref{eq:Learning1WithCoeff}.

\section{Probability estimates}
\label{sec:appendixprob1}
\subsection{Typical Size of $\|x^k\|_1$}
The typical size of  $\gamma(\X_0) = \max_k \|x^k\|_1$ can be directly derived from the following concentration of measure result.
\begin{Theorem}\label{th:gamma}
Let $x$ be a vector of length $\nsig$, whose entries follow the distribution described in Subsection~\ref{subsec:model}, $x_n=\randsign_n g_n$, $n=1\ldots \nsig$. Then for any $\eps>0$
\begin{align}
\P\big(\|x\|_1 > \nsig p ({\textstyle \sqrt{\frac{2}{\pi}}} +\eps)\big) \leq 2\cdot \exp \left(-\frac{\nsig p \cdot \eps^2}{2+\sqrt2 \cdot \eps}\right). \notag
\end{align}
\end{Theorem}
It follows immediately, using a union bound, that with
\begin{equation}
\gamma:=\nsig p ({\textstyle \sqrt{\frac{2}{\pi}}} +\eps),
\label{eq:DefGammaN}
\end{equation}
we have
\begin{align}\label{eq:gammaprob}
\P\big(\gamma(\X_0) > \gamma \big)\leq 2\natoms \cdot \exp \left(-\frac{p\nsig \eps^2}{2+\sqrt2 \cdot \eps}\right).
\end{align}

\subsection{General Approach to Estimating $\alpha$ and $\beta$}

Now we will estimate the probability that for one index $k$ either a) or b) fails. Denote $\Omega_k$ the event 
\[
\Omega_k := \{\radius_q(\bar{\X}_k) < \alpha\} \cup \{\|\X_k (s^k)^\star\|_q > \beta\},
\]
\ie either a) or b) fails for row $k$. Then $\Omega = \cup_k \Omega_k$ is the undesired event $\{\alpha_q(\X_0) < \alpha\} \cup \{\beta_q(\X_0) > \beta\}$. Using a union bound over the row indices $k$ and conditioning on the size of the set of zero entries $|\bad^k|$ we get, 
\begin{align}
\P\big(\Omega \big) 
\leq \sum_{k,M} & \P\big(\Omega_k\ \big|\ |\bad^k|=M \big) \cdot \notag  \P\big( |\bad^k|=M \big) \notag \\
\leq \natoms \cdot& \max_{M \in [M_l, M_u]} \P\big(\Omega^k\ \big|\ |\bad^k|=M \big)\notag\\
& +  \natoms \cdot \P\big( |\bad^k|\notin  [M_l, M_u]  \big)\label{eq:SplitProbSum}.
\end{align}
We start with the estimate of the second term in the sum above, the probability of the number of zero coefficients in a given row being below $M_l$ or above $M_u$.
\begin{lemma}\label{le:SizeGoodBad}
Consider $0< \eps < 1$. Setting $M_l=N(1-p)(1-\eps)$ and $M_u=N(1-p)(1+\eps)$ we get that
\begin{align}
\P\big( |\bad^k|\notin  [M_l, M_u]  \big) \leq 2 \exp(-2N (1-p)^2\eps^2).\label{eq:BoundProbNumZeroCoeff}
\end{align}
\end{lemma}
We will estimate the first term in~\eqref{eq:SplitProbSum} by splitting it into two terms that we will estimate separately
\begin{align}
\P\big(\Omega_k \ \big|\ |\bad^k|=M \big) 
\leq &\ \P\big(\radius_q(\bar{X}_k) < \alpha \ \big|\ |\bad^k|=M \big)\notag \\
&+ \P\big(\|X_k(s^k)^\star\|_q > \beta \ \big|\ |\bad^k|=M \big).
\label{eq:SplitProbSumAlphaBetaGamma}
\end{align}


\subsection{Typical Size of $\alpha_q(\X_0)$}

Now we estimate the typical size of the largest $\ell_q$ ball we can inscribe into the image of the unit cube $Q^{|\bad^k|}$ by $\bar{\X}_k$ when $|\bad^k|=M$. For simplicity we write $L$ for $K-1$, and we denote $A = \bar{X}_k$. From Lemma~\ref{le:radius} we know that we need to estimate the value of $\|A^\star z\|_1$ and compare it to $\|z\|_1$. We begin with some geometrical observations. 
\begin{lemma}\label{le:epscover}
Let $\epscover = \{z_i\}$ be a finite $\eps_\epscover$-cover for the unit $\ell_{q'}$ sphere in $\R^L$. Assume that we have
both the lower bound 
\[
\|A^\star z_i\|_1 \geq \alpha,\ \forall z_i \in \epscover;
\]
and the upper bound
\[
\|A^\star\|_{q' \to 1} = \sup_{\|v\|_{q'} \leq 1} \|A^\star v\|_1  \leq \delta.
\] 
Then
$\radius_\infty(A) \geq \alpha -\delta\eps_\epscover$.
\end{lemma}
\begin{Proof} 
By Lemma~\ref{le:radius} we only need to show that for all $z$ with unit $\ell_{q'}$ norm we have $\|A^\star z\|_1 \geq\alpha-\delta\eps_\epscover$. 
By definition of an $\eps_\epscover$-cover, for all $z$ with unit $\ell_{q'}$ norm we can find $z_i \in \epscover$ with $\|z-z_i\|_{q'} \leq\eps_\epscover$. We then have
\begin{align} 
\|A^\star z\|_1
& \geq \|A^\star z_i\|_1-\|A^\star (z-z_i)\|_1\notag\\
&\geq \alpha - \|A^\star\|_{q' \to1} \cdot \|z-z_i\|_{q'} \geq \alpha-\delta\eps_\epscover.\notag
\end{align}
\end{Proof}
We will therefore estimate a (typical) lower bound for the norm $\|A^\star z_i\|_1$, and an upper bound on the operator norm $\|A^\star\|_{q' \to 1}$. We specialize to the case $q = 2$, but other bounds could be derived for other values of $q$.
\begin{lemma}
\label{th:ballprob}
Let $A=(A_1\ldots A_M)$ be a random matrix of size $L \times M$, whose entries follow the distribution described in Subsection~\ref{subsec:model}, $A_{ij}=\randsign_{ij} g_{ij}$, $i=1\ldots L$, $j=1\ldots M$. Let $z\in\R^L$ be a vector with $\|z\|_2 = 1$. We have the concentration bounds, for $\eps  > 0$,
\begin{align}
\P\big( \|A^\star\|_{2 \to 1}  > M\sqrt{pL}(1 + \eps)\big)
&\leq 2 \exp\left( - \frac{M p \cdot \eps^2}{2+ \sqrt 2 \cdot \eps} \right)\notag.\\
\P\big(\|A^\star z\|_1 \leq Mp( {\textstyle\sqrt{\frac{2}{\pi}}} - \eps)\big) 
& \leq 
2 \exp \left(-\frac{ Mp \cdot \eps^2}{2+\sqrt{2} \cdot \eps}\right).
\end{align}
\end{lemma}
%
%
Combining the above estimates we obtain
\begin{corollary} \label{cor:prob} 
\label{le:SizeAlpha}
Let $0<\eps<1$ and define
\begin{equation}
\label{eq:DefAlphaN}
\alpha := \nsig p(1-p)(1-\eps)({\textstyle \sqrt{\frac{2}{\pi}}}-2\eps - \eps^2)
\end{equation}
Then, for all $M  \in [M_l\ M_u]$ we have
\begin{align}\label{eq:alphaprob}
& \P\big( \radius_2(\bar{\X}_k) < \alpha \ \big|\ |\bad^k|=M\big)  \notag\\
\leq & 2 \cdot \left[\left(\frac{3}{\eps}  \sqrt{\frac{\natoms}{p} }\right)^{\natoms} +1 \right] \cdot \exp \left(-\frac{\nsig p(1-p)(1-\eps)  \eps^2 }{2+\sqrt{2}\eps}\right) 
\end{align}
\end{corollary}

\begin{Proof}
Given $\eps_\epscover \in (0,1)$, we can choose an $\eps_\epscover$-cover $\epscover = \{z_i\}$ for the unit $\ell_2$ sphere in $\R^L$ with $|\epscover| \leq (3/\eps_\epscover)^L$. For a random  $L \times M$ matrix $A=(A_1\ldots A_M)$ distributed as in Lemma~\ref{th:ballprob} we have, combining Lemma~\ref{le:epscover} with Lemma~\ref{th:ballprob} and using $\tilde{\alpha} := Mp(\textstyle \sqrt \frac 2\pi -\eps)$ and $\delta := M \sqrt{pL} (1+\eps)$,
\begin{align}
\P\big(\radius_2(A)  &< \tilde{\alpha}-\delta \eps_\epscover\big) \notag\\
&  \leq \ 
\sum_{z_i \in \epscover} P\big(\|A^\star z_i\|_1\leq \alpha\big)
 +\P\big(\|A^\star\|_{2\to 1} \geq \delta\big).\notag\\
& \leq [(3/\eps_\epscover)^L+1] \cdot  2 \exp \left(-\frac{ Mp \cdot \eps^2}{2+\sqrt{2} \cdot \eps}\right)\notag
\end{align}
Setting $\eps_\epscover=\eps \sqrt{p/L}$ yields
$\tilde{\alpha}-\delta \eps_\epscover = Mp({\textstyle \sqrt{\frac{2}{\pi}}}-2\eps -\eps^2)$.
According to the probability split in~\eqref{eq:SplitProbSum}, we need to find the maximum of the above expression for $M\in [M_l,M_u]$ which is achieved at $M=M_l=\nsig(1-p)(1-\eps)$.

\end{Proof}

\subsection{Typical Size of $\|\X_k(s^k)^\star\|_q$}
We now estimate the size of $\|\X_k(s^k)^\star\|_q$. We need the following theorem.\begin{Theorem}\label{th:beta}
Let $B$ be a random matrix of size $L \times n$, whose entries follow the distribution described in Subsection~\ref{subsec:model}, $B_{ij}=\randsign_{ij} g_{ij}$, $i=1\ldots L$, $j=1\ldots n$, and $s$ be a vector of length $n$ with entries $s_j=\pm 1$, $j=1\ldots n$. Then for $\eps'>0$
\begin{align}
\P\big(\|Bs\|_2^2 \geq Lnp(1+\eps')\big) \leq 2 \exp\left(-\frac{Lp(\eps')^2}{6+2\eps'}\right). \notag
\end{align}
\end{Theorem}
Applying this to the situation at hand, inserting $L=\natoms-1$ and the worst case value for $n=N-M_l=N(p + \eps - \eps p)$ and setting $ \eps' = (\nsig/L) \eps$ we get:
\begin{lemma}\label{le:SizeBeta}
Define
\begin{equation}
\beta:= \nsig p \sqrt{\textstyle (\frac{K-1}{\nsig}+\eps)(1+\frac{\eps}{p}-\eps )},
\label{eq:DefBetaN}
\end{equation}
For any $M \in [M_l, M_u]$ we have
\begin{align}\label{eq:betaprob}
\P\big(&\|X_k(s^k)^\star\|_2 > \beta \big| |\bad^k|=M \big) \notag\\
\leq & 2 \cdot \exp\left(-\frac{\nsig p\eps^2}{6{\textstyle \frac{K-1}{\nsig}}+2\eps}\right).
\end{align}
\end{lemma}


\section{
Concentration Inequalities}\label{app:probability_estimates}
Here we will sketch the proofs of the concentration inequalities used in the previous section. They are based on a special version of Bernstein's inequality, see e.g. \cite{bennett62}.
\begin{Theorem} Let $Y_i$, $i=1\ldots M$, be independent random variables with 
\begin{align}
\E(Y^2_i)\leq v^2 \qquad \mbox{ and } \qquad \E(|Y_i|^k )\leq \frac{1}{2} k!\,v^2 c^{k-2},\ k > 2. \label{decaycond}
\end{align} Then
\begin{align}
\P\Big(|\sum_{i=1}^M (Y_i - \E(Y_i))|>\eps\Big) \leq 2 \exp\Big( - \frac{\eps^2}{2(M v^2 +c \eps)} \Big).\notag
\end{align}
\end{Theorem}
We will also use Hoeffding's inequality.
\begin{Theorem}[Hoeffding's inequality]
Let $Y_1\hdots Y_\nsig$ be independent random variables. Assume that the $Y_n$ are almost surely bounded, meaning for $1\leq i\leq \nsig$ we have $\P(Y_n\in [a_n,b_n])=1$. Then, for the sum of these variables $S=Y_1+\ldots+Y_\nsig$ we have the inequality
\begin{align}
\P(S-\E(S) \geq \nsig t)\leq \exp(-\frac{2\nsig^2t^2}{\sum^\nsig_{n=1} (b_n-a_n)^2}),\notag
\end{align}
which is valid for positive values of $t$. $\E(S)$ is the expected value of $S$.
\end{Theorem}

\subsection{Proof of Lemma~\ref{le:SizeGoodBad}}
In each row of $\X$, the number of zero coefficients $|\bad^k|$ is $\nsig$ minus the number of non-zero coefficients $|\good^k|$, which is the sum of the indicator variables $\sum_n \randsign_{kn}$. The $\randsign_{nk}$ are taking only the values zero and one, so we can use Hoeffding's inequality with $a_i=0$, $b_i=1$ and $\E(\sum_n \randsign_{kn})=p\nsig$, leading to 
\begin{align}
\P(|\good^k|-p\nsig \geq \nsig t)\leq \exp(-2N t^2).\notag
\end{align}
Choosing $t=(1-p)\eps$ and 
using $|\bad^k|=\nsig-|\good^k|$ we get
\begin{align}
\P(|\bad^k| \leq N(1-p)(1-\eps) )\leq \exp(-2N (1-p)^2\eps^2).\notag
\end{align}
To bound the converse probability that $|\bad^k|$ is very large, we set $Y_n = 1-\randsign_{kn}$ and again $t=(1-p)\eps$ to get directly to 
\begin{align}
\P(|\bad^k| \geq N(1-p)(1+ \eps) )\leq \exp(-2N (1-p)^2\eps^2).\notag
\end{align}

\subsection{Proof of Theorem~\ref{th:gamma}}
Since $\|x\|_1 = \sum_{i=1}^N \randsign_i |g_i|$, we will use the Bernstein inequality with $Y_i=  \randsign_i \cdot |g_i|$. The moments of $\randsign_i$ are constant equal to $p$. The random variable $|g_i|$ follows a Chi-distribution of degree 1 so its moments are
\begin{equation}
\label{eq:MomentsChiDistribution}
\E(|g_i|^k)=2^{\frac{k}{2}} \frac{\Gamma(\frac{k+1}{2})}{\Gamma(\frac{1}{2})}
\end{equation}
Especially, we have $\E(Y_i)=p\sqrt{\frac{2}{\pi}}$ and $\E(|Y_i|^2)=p$, and using the recurrence relation for the Gamma function $\Gamma(t+1)=t\Gamma(t)$ and $\sqrt{2}/\Gamma(\frac{1}{2})=\sqrt{\frac{2}{\pi}}<1$ we can bound by induction the moments of $Y_i$ for $k\geq2$ as
\begin{equation}
\label{eq:MomentsYDistribution}
\E(|Y_i|^k) \leq p\cdot \frac{k!}{ 2^{k/2}},\ k \geq 2,
\end{equation}
so the moments suffice Condition~\eqref{decaycond} with $c=1/\sqrt{2}$ and we get
\begin{align}
\P(\|x\|_1 > Np \textstyle\sqrt{\frac{2}{\pi}}  + \eps)
&\leq 2 \cdot \exp\left( - \frac{\eps^2}{2(Mp + \eps/\sqrt{2})} \right)\notag.
\end{align}
Setting $\eps = Mp \cdot \eps'$ yields the result.
\subsection{Proof of Lemma~\ref{th:ballprob} -- first part}

To bound $\|A^\star\|_{2 \to 1}$ we begin by using the crude bound $\|A^\star\|_{2 \to 1} = \|A\|_{1 \to 2} \leq \sum_{i=1}^M\|A_i\|_2$.
We set $Y_i=\|A_i\|_2 =( \sum_{j=1}^L \randsign_{ij}^2g_{ij}^2)^\frac{1}{2}$. All $Y_i$ are identically distributed so for the analysis we can drop the subscript $i$. We can calculate directly 
\begin{align}
\E(Y^2)=\E( \sum_{j=1}^L \randsign_{j}^2g_{j}^2)=pL.\notag
\end{align}
For the higher order moments $k>2$ we use a little trick to separate the expectation over $\randsign$ and $g$,
\begin{align}
\E Y^k&=\E_g\E_\randsign \big( \sum_{j=1}^n \randsign_{j}^2g_{j}^2\big)^\frac{k}{2}=\E_g \Big( \big(\sum g_j^2\big)^\frac{k}{2} \E_\randsign \big( \frac{ \sum \randsign_{j}^2g_{j}^2}{\sum g_j^2}\big)^\frac{k}{2}\Big).\notag
\end{align}
The fraction in the last expression is always smaller than 1 so for $k>2$ we have
\begin{align}
\E Y^k &\leq\E_g \Big(\big(\sum g_j^2\big)^\frac{k}{2} \E_\randsign \big( \frac{ \sum \randsign_{j}^2g_{j}^2}{\sum g_j^2}\big)\Big)=p\cdot \notag \E_g \left( \big(\sum g_j^2\big)^\frac{k}{2}\right).
\end{align}
The random variable $\tilde{Y}=\big(\sum_{j=1}^n g_j^2\big)^\frac{1}{2}$ follows a Chi-distribution of degree $L$ so for its $k$-th moments we have the formula
\begin{align}
 \E(\tilde{Y}^k)=2^{\frac{k}{2}} \frac{\Gamma(\frac{k+L}{2})}{\Gamma(\frac{L}{2})}.\notag
\end{align}
A long and tedious calculation involving the recurrence formula for the Gamma function, Stirling's formula and treating both cases, $k$ is even respectively odd, yields the bound
$  \E(\tilde{Y}^k)\leq (\frac{L}{2})^{k/2} k!$. This leads to $\E(Y^k)\leq p \frac{L}{2}^{k/2} k$, meaning that the higher order moments follow the decay condition in~\eqref{decaycond} for $c=\sqrt{L/2}$.
Together with the following bound for the first order moment,
\begin{align}
 \E(Y)\leq \E(Y^2)^{\frac{1}{2}} = \sqrt{pL},\notag
\end{align}
we get
\begin{align}
\P\big( \|A^\star\|_{2 \to 1} 
> M\sqrt{pL} + \eps\big)
&\leq 2 \exp\Big( - \frac{\eps^2}{2(MpL +\eps \sqrt{L/2})} \Big)\notag.
\end{align}
To get the version of the formula used in Section~\ref{sec:probana} simply set $\eps=M\sqrt{pL} \cdot \eps'$ and observe that since $p < 1$
\[
\frac{\eps^2}{2(MpL +\eps \sqrt{L/2})} 
=
\frac{M\sqrt{p}(\eps')^2}{2\sqrt{p} +\sqrt{2} \eps'} 
\geq
\frac{Mp(\eps')^2}{2 +\sqrt{2} \eps'} 
\]

\subsection{Proof of Lemma~\ref{th:ballprob} -- second part}

To lower bound $\|A^\star z\|_1$ we expand it as 
\begin{align}
\|A^\star z\|_1= \sum_{i=1}^M |\ip{A_i}{z}| = \sum_{i=1}^M |\sum_{j=1}^n \randsign_{ij}g_{ij}z_j|:= \sum_{i=1}^M Y_i. \notag
\end{align}
The random variables $Y_i$ all follow the same distribution so it suffices to calculate the moments of $Y= |\sum_{j=1}^n \randsign_{j}g_{j}z_j|$. Define $\tilde{Y} =\sum_{j=1}^n \randsign_{j}g_{j}z_j$. Since the $g_k$ are \iid zero mean Gaussians with variance $\sigma^2= 1$, $\tilde{Y}$ is zero mean Gaussian with variance $\tilde{\sigma}^2= \sum_{j=1}^n z_j^2\randsign^2_{j}:=\|z\randsign\|_2^2$ and we get
\begin{align}
\E(|Y|^k)=\E(|\tilde{Y}|^k|)=\E(|\|z\randsign\|_2 \cdot g_1|^k)=\E_\randsign(\|z\randsign\|_2^k) \cdot \E_g(|g_1|^k)
\end{align}
Since $\|z\randsign\|_2 \leq \|z\|_2 = 1$, we have for $k\geq 2$ 
\begin{align}
\E_\randsign(\|z\randsign\|_2^k) \leq \E_\randsign(\|z\randsign\|_2^2) = \E_\randsign\big(\sum_{j=1}^n z_j^2\randsign^2_{j}\big) \leq p,\notag
\end{align}
while for $k=1$ we get
\begin{align}
\E_\randsign(\|z\randsign\|_2)= \E_\randsign\Big(\big(\sum_{j=1}^n z_j^2\randsign^2_{j}\big)^\frac{1}{2}\Big)\geq \E_\randsign\left(\sum_{j=1}^n z_j^2\randsign^2_{j}\right)=p.\notag
\end{align}
Again, $|g_1|$ is Chi-distributed of degree 1 so its moments are given by~\eqref{eq:MomentsChiDistribution} and the moments of $Y_i$ are thus bounded by~\eqref{eq:MomentsYDistribution}, which suffices the decay condition in~\eqref{decaycond} for $c=1/\sqrt{2}$. As a result
\begin{align}
\P\Big(\|A^\star z\|_1 < M\E(|Y|)- \eps\Big) 
&< 2 \exp\left( - \frac{\eps^2}{2(Mp +\eps/\sqrt{2})} \right). \notag
\end{align}
Together with the bound for $\E(|Y|)\geq p\sqrt{\frac{2}{\pi}}$, setting $\eps= Mp \cdot \eps'$ leads to the final form of the bound used in Section~\ref{sec:probana}.

\subsection{Proof of Theorem~\ref{th:beta}}
We expand $\|Bs\|^2_2=\sum_{i=1}^L |\ip{B^i}{s}|^2$, where $B^i$ denotes the $i$-th row of $B$. and set
$Y_i = |\ip{B^i}{s}|^2=(\sum_{j=1}^n \randsign_{ij}g_{ij} s_j )^2$. Since the $Y_i$ are again identically distributed we drop the subscript $i$ for the analysis. First we get,
\begin{align}
\E(Y)=\E\Big( \big(\sum_{j=1}^n \randsign_{j}g_{j} s_j \big)^2\Big)=\E\Big( \sum_{j=1}^n \randsign^2_{j}g^2_{j} s^2_j \Big)=p \cdot n .\notag
\end{align}
Observe that $\sum \randsign_{j}g_{j} s_j$ is again Gaussian and distributed like $(\sum \randsign^2_{j} s^2_j)^\frac{1}{2} \cdot g_1 = \|\randsign\|_2 \cdot g_1$. Hence,
\begin{align}
\E(Y^k)=\E\Big( \big(\sum_{j=1}^n \randsign_{j}g_{j} s_j \big)^{2k}\Big)&=\E_\randsign \E_g (\|\randsign\|_2^{2k} g_1^{2k} )\notag\\
&= \E_\randsign(\|\randsign\|_2^{2k} )  \E_g (g_1^{2k} ). \notag
\end{align}
For the even Gaussian moments we have the formula $\E_g (g_1^{2k})=\frac{(2k)!}{2^k k!}$, while the term depending on $\randsign$ can be bounded as
\begin{align}
\E_\randsign(\|\randsign\|_2^{2k} )& = \E_\randsign\Big( \big(\sum_{j=1}^n \randsign_j^2\big)^k \Big)= n^k \cdot   \E_\randsign\Big( \big(\frac{1}{n}\sum_{j=1}^n \randsign_j^2\big)^k \Big) \notag \\
&\leq n^k \cdot   \E_\randsign\Big( \frac{1}{n}\sum_{j=1}^n \randsign_j^2 \Big) =n^k \cdot p,\notag
\end{align}
leading to $E(Y^k)\leq p n^k \frac{(2k)!}{2^k k!}$. Especially for $k=2$ we have $E(Y^2)\leq 3p n^2$ and so for $k>2$ we can estimate 
\begin{align}
E(Y^k)\leq 3pn^2 \frac{1}{3} n^{k-2} \frac{(2k)!}{2^k k!} \leq \ldots \leq \frac{1}{2} \E(Y^2) (2n)^{k-2} k!, \notag
\end{align}
meaning that the moments follow the decay condition in~\eqref{decaycond} with $c=2n$ and therefore
\begin{align}
\P\left(\|Bs\|^2_2 > Lnp + \eps\right)
&\leq 2 \exp\left( - \frac{\eps^2}{6pn^2L + 2n\eps} \right)\notag.
\end{align}
Again setting $\eps= Lnp \cdot \eps'$ leads to the final version.

\section{Proof of Main Theorem}\label{app:proof_mth}

First, we observe that if $p\leq4/5$ and $\natoms/\nsig \leq 1/3$ all the appearing exponentials can be upper bounded by 
\begin{align}
\exp\Big( - Np(1-p) \frac{\eps^2(1-2\eps)}{2}\Big).\notag
\end{align}
Therefore, with the definition of $\alpha,\beta,\gamma$ in~ \eqref{eq:DefAlphaN}, \eqref{eq:DefBetaN} and \eqref{eq:DefGammaN} we obtain from Lemmata~\ref{le:SizeAlpha},~\ref{le:SizeBeta},~\ref{th:gamma} that we have 
\[
\frac{\alpha_2(\X_0)-\beta_2(\X_0)}{\gamma(\X_0)} \geq \frac{\alpha-\beta}{\gamma}
\]
except with probability at most
\begin{align}
2K &\left[ \left(\frac{3}{\eps}  \sqrt{\textstyle\frac{\natoms}{p} } \right)^{\natoms} +3\right] \cdot \exp \Big( - Np(1-p) \frac{\eps^2(1-2\eps)}{2}\Big)\notag\\
& \leq 4K \left(\frac{3}{\eps}  \sqrt{\textstyle\frac{\natoms}{p} } \right)^{\natoms} \cdot \exp \Big( - Np(1-p) \frac{\eps^2(1-2\eps)}{2}\Big) \notag\\
& = 4K \exp\Big( \textstyle\frac{\natoms}{2} \log\left(\frac{9\natoms}{\eps^2 p}\right) - Np(1-p) \frac{\eps^2(1-2\eps)}{2}\Big).\label{eq:FinalProbBound}
\end{align}
Next, observe that for the right hand side to be smaller than 1, we need that $\eps<1/2$ and $\nsig p(1-p)\eps^2 > \natoms$. Consequently
\begin{align}
\natoms/N<p(1-p)\eps^2<1/16,\notag
\end{align}
meaning that whenever $\natoms/\nsig > 1/3$ the probability bound is trivially true, and we only need to assume $p \leq 4/5$.

Now, from Theorem~\ref{TH:SCRECOVERYEXPLICITCOHERENCE} we know that any sufficiently incoherent basis satisfying $\max_k \|\bar{m}_k\|_2 < (\alpha-\beta)/ \gamma$ will therefore be locally identifiable by $\ell_1$ minimization, except with probability at most equal to the right hand side in~\eqref{eq:FinalProbBound}.

Inserting the values for $\alpha,\beta, \gamma$ from \eqref{eq:DefAlphaN}, \eqref{eq:DefBetaN} and \eqref{eq:DefGammaN} we can lower bound the maximally allowed coherence $(\alpha-\beta)/\gamma$ with
\begin{align}
 &\frac{(1-p)(1-\eps)({\textstyle \sqrt{\frac{2}{\pi}}}-2\eps-\eps^2)-  \sqrt{\textstyle (\frac{\natoms}{\nsig}+\eps)(1+\frac{\eps}{p}-\eps )}
}{ ({\textstyle \sqrt{\frac{2}{\pi}}} +\eps)} \notag \\
&\geq  (1-p) \cdot (1-5 \eps) - \textstyle\sqrt{\frac{\pi}{2}\left(\frac{\natoms}{\nsig}+\eps\right)\left(1+\frac{\eps}{p}\right)}.\notag
\end{align}

\bibliography{/Users/karinschnass/Desktop/latexnotes/karinbibtex.bib}
\bibliographystyle{plain}
\end{document}